\documentclass[reprint, 
 superscriptaddress,
 amsmath,
 amssymb,
 aps,
 pra]{revtex4-2}

\usepackage{xcolor}
\usepackage{graphicx}
\usepackage{dcolumn}
\usepackage{bm}
\usepackage{xr-hyper}
\usepackage{hyperref}
\usepackage{tabularx}

\begin{document}

\title{Optimizing information transmission in neural induction constrains cell surface contacts of ascidian embryos}

\author{Rossana Bettoni}
\affiliation{Unit\'{e} de Chronobiologie Th\'{e}orique, Facult\'{e} des Sciences, CP231, Universit\'{e} Libre de Bruxelles (ULB), Boulevard du Triomphe, Brussels, Belgium}
\affiliation{Applied Physics Research Group, Vrije Universiteit Brussel, Brussels, Belgium}
\affiliation{Interuniversity Institute of Bioinformatics in Brussels, ULB-VUB, La Plaine Campus, Brussels, Belgium}

\author{Genevi\`{e}ve Dupont}
\affiliation{Unit\'{e} de Chronobiologie Th\'{e}orique, Facult\'{e} des Sciences, CP231, Universit\'{e} Libre de Bruxelles (ULB), Boulevard du Triomphe, Brussels, Belgium}
\affiliation{Interuniversity Institute of Bioinformatics in Brussels, ULB-VUB, La Plaine Campus, Brussels, Belgium}

\author{Aleksandra M. Walczak}
\altaffiliation{Co-last authors}
\affiliation{Laboratoire de physique de l'\'{E}cole normale sup\'{e}rieure, CNRS, Paris Sciences et Lettres (PSL) University, Sorbonne Universit\'{e}, and Universit\'{e} de Paris, 75005 Paris, France.}
\affiliation{Applied Physics Research Group, Vrije Universiteit Brussel, Brussels, Belgium}

\author{Sophie de Buyl}
\altaffiliation{Co-last authors}
\affiliation{Applied Physics Research Group, Vrije Universiteit Brussel, Brussels, Belgium}

\affiliation{Interuniversity Institute of Bioinformatics in Brussels, ULB-VUB, La Plaine Campus, Brussels, Belgium}
\affiliation{Data Analytics Laboratory, Vrije Universiteit Brussel, Brussels, Belgium}

\date{October 22, 2024}

\begin{abstract}
The onset of neural induction in the anterior ectoderm of ascidian embryos is regulated at the extracellular level by FGF signaling molecules, which control the acquisition of  neural fate through the activation of the ERK pathway. Among the anterior ectoderm cells exposed to FGF, only a fraction will acquire neural fate. The selection of neural precursors depends on the quasi-invariant geometry of the embryo, which imposes upon each ectoderm cell a precise area of cell surface contact with underlying FGF-expressing (mesendoderm) cells.
Here, we investigate information transmission between FGF and activated ERK and how this depends on the geometry of the system. Optimizing information transmission with the constraint that the total FGF-emitting surface area is restricted, as in the embryo, we find that the surface contacts with FGF that maximize information transmission are close to those observed experimentally. 
This information optimal solution is compatible with the anterior ectoderm cells having different areas of cell surface exposure to FGF, allowing the embryo to use cell surface areas as a regulatory mechanism for differentiating the outcome of cells that sense a constant FGF concentration.
\end{abstract}

\maketitle

\section{Introduction}

The information contained in the fertilized egg must be decoded by the developing embryo to transform the latter into an organized structure.
This is achieved by combining mechanical interactions constrained by geometry and molecular signaling that is stochastic by nature, thereby limiting the transmission of the information.  
Optimization principles have been suggested to assess how close to the maximal possible transmitted information biological systems can operate \cite{laughlinSimpleCodingProcedure1981, tkacikInformationFlowOptimization2008, petkovaOptimalDecodingCellular2019, brucknerInformationContentOptimization2023}. Specifically, inspired by developmental regulation in early fly embryos, optimizing information between a spatially decaying input morphogen and downstream gene readouts has led to tiling solutions, where different genes make a readout in non-overlapping concentration domains \cite{tkacikOptimizingInformationFlow2009, tkacikInformationTransmissionGenetic2011}. While the real regulatory networks in early fly development have proven more complex~\cite{petkovaOptimalDecodingCellular2019, Fernandes2022},  optimization of information transmission can contribute to explain  
the structure of cellular and molecular interactions that lead to the reproducibility of development and the emergence of a highly organized organism from a fertilized egg~\cite{tkacikManyBitsPositional2021, brucknerInformationContentOptimization2023}. Development is driven by the coordination and mutual dependence of genetically encoded signals, mechanical stress and geometry~\cite{Miller2013}. However while inference approaches have been proposed to link mechanical stresses and lineage structure in development~\cite{liuMechanicalAtlasAscidian2022, ichbiahEmbryoMechanicsCartography2023}, optimizing information transmission of signals has not been linked to embryo geometry. Here we use the well studied case of ascidian embryos to explore how information transmission of concentration-dependent signals constrains cell geometry in early embryos.

Ascidians exhibit a very simple and reproducible embryogenesis. Of particular interest is the fact that they develop, like nematodes, with an invariant cleavage pattern. This means that the positioning and timing of cell divisions are quasi-invariant between different individuals of the same species  \cite{dumollardInvariantCleavagePattern2017}, leading to an extreme form of canalization \cite{waddingtonCANALIZATIONDEVELOPMENTINHERITANCE1942}. Moreover, when the embryo consists of only 112-cells, the segregation of the major tissue-specific cell lineages (muscle, endoderm, nervous system etc) is almost completed. There is also no embryo growth or cell migration during early ascidian development. This type of embryogenesis makes it possible to follow each cell individually and trace the establishment of their fates, and therefore to address the question of cell fate segregation and the establishment of the body plan in a relatively simple context. 
Additionally, ascidian embryogenesis is extremely well documented
\cite{lemaireEvolutionaryCrossroadsDevelopmental2011}. Databases containing information about the geometry of the embryo, gene regulatory networks leading to the larval body plan and cell specific gene expression patterns are publicly available \cite{brozovicANISEED2017Extending2018,imaiRegulatoryBlueprintChordate2006,satouGeneRegulatorySystems2015,sharmaSinglecellTranscriptomeProfiling2019,guignardContactAreadependentCell2020, winkleySinglecellAnalysisCell2021}. 
Recent works inferred cell pressures, membrane and line tensions at single-cell resolution reproducing the mechanical cell stresses observed during the early stages of ascidian gastrulation ~\cite{liuMechanicalAtlasAscidian2022,ichbiahEmbryoMechanicsCartography2023}. Altogether these properties make ascidian embryogenesis an appealing toy model to study cell fate decisions.

\begin{figure*}[ht]
    \centering
    \includegraphics[width=1\textwidth]{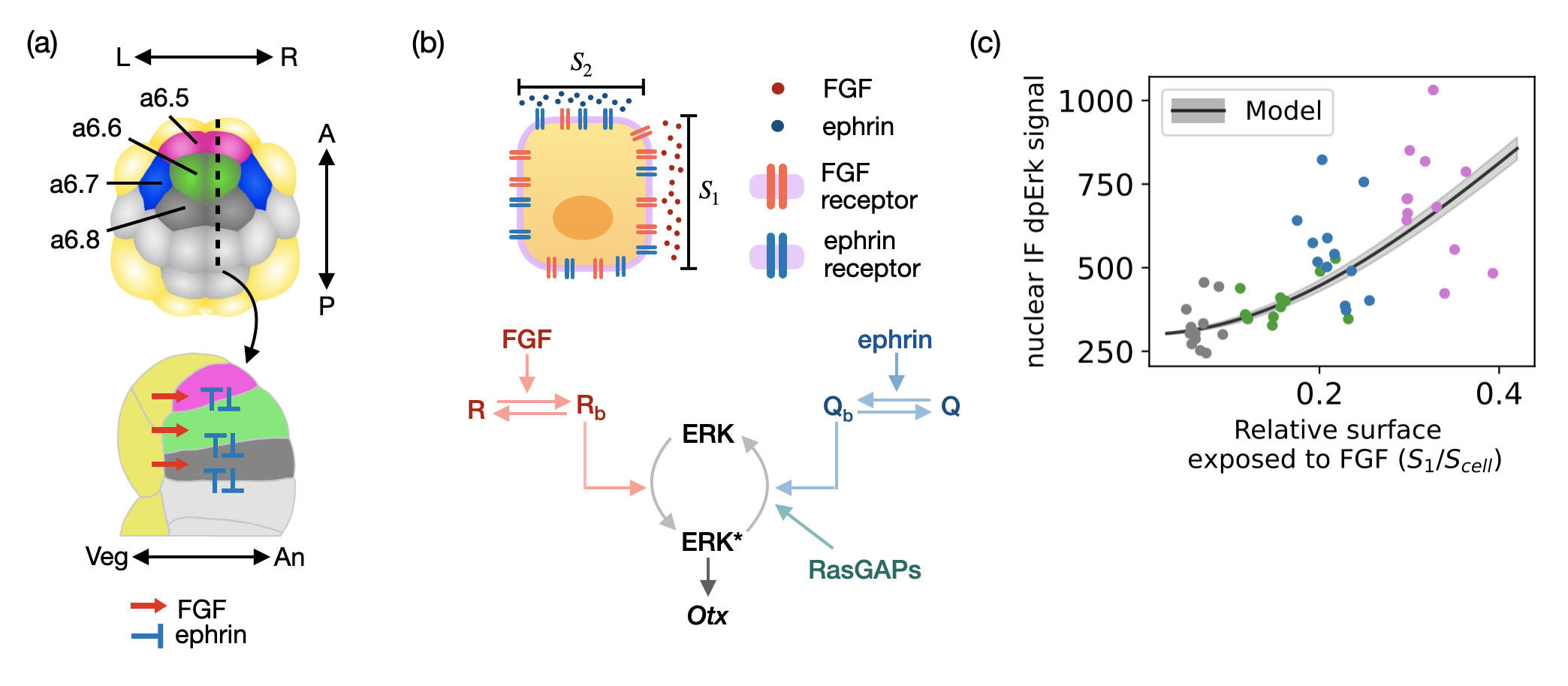}
    \caption{(a) Cartoon of the 32-cell stage ascidian embryo. View of the animal side of the embryo (top). Anterior ectoderm cells (a-line cells), which can become neural, are indicated in different colors: a6.5 cells in magenta, a6.6 cells in green, a6.7 cells in blue, and a6.8 cells in dark gray. Mesendoderm cells are shown in yellow and posterior ectoderm cells (b-line cells) in light gray. The black dotted line is used for the sagittal section of the embryo shown at the bottom. FGF and ephrin signals are represented by red and blue arrows, respectively. The anterior-posterior (A-P), left-right (L-R) and vegetal-animal (Veg-An) axis of the embryo are indicated by double-sided arrows. The total cell surface areas of the different a-line cells are: $S_{\rm cell}^{a6.5} = 6837 \pm 248\ \mu m^2$,  $S_{\rm cell}^{a6.6}= 8268 \pm 172 \ \mu m^2$, $S_{\rm cell}^{a6.7}= 7331 \pm 310 \ \mu m^2$, $S_{\rm cell}^{a6.8}= 8646 \pm 135 \ \mu m^2$. Figure adapted from \cite{williaumeCellGeometrySignal2021}. (b) Cartoon of the pathway leading to neural induction. The process is initiated at the extracellular level by the binding of FGF and ephrin to their receptors ($R$ and $Q$, respectively) present on the cell membrane. The areas of the cell surface exposed to FGF and ephrin ligands are denoted by $S_1$ and $S_2$. For simplicity the membrane bound ephrin ligands are not shown attached to the membrane of the emitting cell.  Bound FGF receptors ($R_b$) induce the double phosphorylation of ERK. The active form of ERK (dpERK) is indicated in the scheme as ERK*. Bound ephrin receptors ($Q_b$) favor the de-activation of ERK*, which is also promoted by ephrin-independent RasGAPs molecules (indicated as RasGAPs in the scheme). ERK* induces the expression of the neural marker $Otx$. (c) Nuclear dpERK immunofluorescence (IF) signals measured experimentally in individual a-line cells as a function of the relative area of cell surface contact with FGF-expressing cells ($S_1/S_{\rm cell}$). Experimental data obtained in the different cell types are shown with dots of different colors: a6.8 cells are shown in gray, a6.6 cells in green, a6.7 cells in blue, and a6.5 cells in magenta. The experimental data are compared with the model predictions (black line).  The shaded region represents the noise in the level of ERK* fluorescence predicted by the model using the Langevin approach.
    The model predictions are obtained as described in section \ref{sec_model}, \ref{Langevin approach} and \ref{reproduction_expData}, with $c=5$.
    The parameter values used to reproduce the experimental data are listed in Table~\ref{tab1}. $A = 8000$, $B=280$, see section \ref{reproduction_expData}.}
    \label{F1}
\end{figure*}

Contrary to the role that morphogen gradients have been shown to play during patterning of many other systems \cite{christianMorphogenGradientsDevelopment2012}, in ascidians, short-range signals between contacting cells could explain all known early embryonic inductions \cite{guignardContactAreadependentCell2020}. In support of this, in our previous work we have shown that, during early neural induction at the 32-cell stage of development, measured areas of cell surface contact act as strong predictive determinants for differential cell fate specification \cite{bettoniModelNeuralInduction2023, williaumeCellGeometrySignal2021}. During neural induction, in response to an extracellular inductive signal (i.e., the fibroblast growth factor), specific cells in the embryo express the gene $Otx$. The process is mediated by the ERK signaling pathway (see section \ref{NI} for a detailed description of the neural induction process).
Building on our previous studies, we analyze information transmission in the first step of neural induction: from the inducer FGF to the activation of ERK.  We investigate whether the contact surfaces of four pairs of cells (a-line cells, Fig. \ref{F1}(a)) involved in the process of neural induction are compatible with optimal information transmission between the inductive signal and ERK. More precisely, we ask if the probability distribution of the concentration of the inducer, i.e., the fibroblast growth factor (FGF), is read optimally by the cells through the portion of their surfaces that is in contact with this ligand. 

To address this question we develop, in section \ref{sec_model}, a stochastic model of the pathway involved in neural induction. This is a minimal version of the deterministic model we formulated in \cite{bettoniModelNeuralInduction2023, williaumeCellGeometrySignal2021}. In section \ref{sec_one-cell}, relying on the stochastic minimal model, we analyze information transmission in a single cell. More precisely, we investigate whether the information transmitted between FGF and the number of active ERK molecules (ERK*) depends on the geometrical configuration of the cell, i.e. on the surface of the cell in contact with FGF. 
In section \ref{symmetric_case},  we investigate how the geometry of the system influences information transmission in the multi-cellular context. In particular, we consider the case in which all the cells are identical, except for their surface contacts with FGF. We perform this analysis in the presence of an external constraint which limits the total area of the cell surface contacts with FGF. 
To formulate this constraint problem, we first consider two cells and then we generalize the computation to the four-cell case. 
Finally, in section \ref{sec_four-cells} we turn to the case relevant for ascidian embryos and consider four cells with different total surfaces. We compute the areas of the cell surface contacts with FGF that maximize information transmission and compare them with the experimental measurements. 

\section{Neural induction in the ascidian embryo \label{NI}}

In ascidians, neural induction in anterior ectoderm cells is the first step of a complex process of cell fate specification that will generate the anterior-most part of the larval central nervous system together with anterior placode-like structures  \cite{lemaireEarlyStepsFormation2002,liuGeneticProgramSpecify2020}.
It takes place at the 32-cell stage of embryonic development (schematized in Fig.~\ref{F1}(a)), during a single cell cycle of about 30 minutes \cite{williaumeCellGeometrySignal2021}. 
The embryo exhibits left-right symmetry and is  divided into animal and vegetal hemispheres. The animal hemisphere will eventually give rise to mostly ectodermal tissues,  the epidermis and part of the nervous system. The vegetal hemisphere generates mostly mesendoderm fates, the endoderm and mesoderm tissues including notochord and muscle as well as posterior central nervous system  \cite{guignardContactAreadependentCell2020}. 

During neural induction, 2 out of 8 anterior ectoderm cells (a-line cells, shown in magenta, green, blue and dark gray in Fig. \ref{F1}(a))  transcriptionally activate the neural marker $Otx$ \cite{hudsonInductionAnteriorNeural2001, williaumeCellGeometrySignal2021}. The signaling pathway that controls the expression of $Otx$ is complex, involving  a large number of reactions \cite{bettoniModelNeuralInduction2023, lavoieERKSignallingMaster2020}. A simplified scheme of the pathway is shown in Fig. \ref{F1}(b)). The process is initiated at the extracellular level by two signaling molecules: FGF and ephrin. FGF binding to its receptors $R$ on the cell membrane activates the ERK signaling pathway which in turn induces the expression of $Otx$. At the same time, ephrin binding to its receptors $Q$ on the cell membrane attenuates the activation of the ERK pathway thereby reducing the level of $Otx$ expression. The activation of ERK (and therefore the expression of $Otx$) is also inhibited by ephrin-independent RasGAPs molecules (indicated as RasGAPs in  Fig. \ref{F1}(b)). 

All the a-line cells are assumed to perceive 
the same concentration of FGF,  produced by the underlying mesendoderm cells (shown in yellow in Fig. \ref{F1}(a)) \cite{williaumeCellGeometrySignal2021}. At the same time, the a-line cells are also in contact with ephrin,  produced by all ectoderm cells. Despite the observation that all a-line cells are competent for neural induction and all of them are in contact with FGF, only the a6.5 pair of cells will undergo neural induction. Selection of neural precursors is controlled by the geometry of ectoderm cells, which  dictates their contact surface areas  with their   FGF- and ephrin-expressing neighbors \cite{bettoniModelNeuralInduction2023, williaumeCellGeometrySignal2021}.

\section{Results }

\subsection{Stochastic minimal model of ERK activation during neural induction \label{sec_model}}

In our previous work ~\cite{bettoniModelNeuralInduction2023, williaumeCellGeometrySignal2021} we developed a computational model to describe the regulation of $Otx$ expression by the FGF- and ephrin-regulated ERK pathway during ascidian neural induction. The model allowed us to show that differences in the surface contacts with FGF and ephrin are sufficient to control the process of neural induction at this stage in the embryo.  

To address the question of  information transmission, we develop a stochastic model of the pathway, reducing our previously developed model \cite{bettoniModelNeuralInduction2023} to facilitate the derivation of analytical results and improve interpretability.  
The model consists of three stochastic differential equations describing the temporal evolution of the number of bound FGF receptors $R_b$, the number of bound ephrin receptors  $Q_b$ and  the number of active (i.e., doubly phosphorylated) ERK molecules ($E^*$): 
\begin{equation}
     \dot{R}_b = k_{d\scriptscriptstyle+}c (R-R_b) - k_{d\scriptscriptstyle-} R_b+ \xi_{R} ;
     \label{LangevinEq_Rb}
\end{equation}
\begin{equation}
 \dot{Q}_b= k_{e\scriptscriptstyle+} e (Q-Q_b) - k_{e\scriptscriptstyle-} Q_b+ \xi_{Q} ;
 \label{LangevinEq_Qb}
\end{equation}
\begin{eqnarray}
\dot{E}^* &= &V_s \frac{R_b^2}{R_b^2 + K_s^2}(E_T- E^*) \label{LangevinEq_E*} \\ \nonumber
&& - \left ( V_{rg}\frac{Q_b}{Q_b + K_{rg}} + k \right ) E^*+ \xi_{E}.
\end{eqnarray}
We consider relative values of FGF ($c=[FGF]/[FGF]_0$) and  ephrin ($e=[{\rm eph}]/[{\rm eph}]_0$) concentrations, where  $[FGF]_0$ and $[{\rm eph}]_0$ are baseline values of FGF and ephrin concentrations ($\simeq 0.1 $ $nM$). $R$ and $Q$ are the number of FGF and ephrin receptors exposed to FGF or ephrin in a cell. 
$k_{d-}$ and $k_{e-}$  are the unbinding rate constants of FGF and ephrin to their receptor. 
$k_{d+}$ and $k_{e+}$ are the binding rate constants of FGF and ephrin to their receptor multiplied by the baseline values $[{\rm FGF}]_0$ and $[{\rm eph}]_0$, respectively. 
$E_T$ is the total number of ERK molecules present inside the cell. $V_s$ and $V_{rg}$ are the maximum rates of ERK activation and deactivation, $K_s$ and $K_{rg}$ are the half saturation constants for $R_b$ and $Q_b$. The parameter $k$ is a deactivation rate constant modeling the presence of ephrin-independent RasGAP activity \cite{williaumeCellGeometrySignal2021}. 
$\xi_{R}$, $\xi_{Q}$ and $ \xi_E$ denote white noise with zero mean and amplitudes $A_R$,  $A_Q$ and $A_E$ given by Eqns.~\eqref{AE}, ~\eqref{An} and ~\eqref{Am} (see section~\ref{Langevin approach}). 
All parameter values are given in Table~\ref{tab1}.

Similarly to our previous model, we make two key assumptions 
\cite{williaumeCellGeometrySignal2021}. First, we assume that the density of FGF and ephrin receptors on the membrane of the cell is uniform. 
This implies that the number of receptors exposed to FGF ($\rm R$) or ephrin ($\rm Q$) in a cell is proportional to the fraction of the cell surface area exposed to FGF ($S_1$) or to ephrin ($S_2$): 
\begin{equation}
    R = R_T\frac{S_1}{S_{\rm cell}}; \quad \quad Q = Q_T\frac{S_2}{S_{\rm cell}};
    \label{receptors}
\end{equation}
where $R_T$ and $Q_T$ are the total number of FGF and ephrin receptors present on a cell and $S_{\rm cell}$ is the total surface of the cell. 
Second, we consider that all the reactions reach steady-state fast compared to the time-scale of the process (which occurs over 30 minutes). This is supported by the observation that the ERK pattern (a6.5$>$a6.7$>$a6.6$>$a6.8) appears to be established rather quickly, short after these cells form. This assumption allows us solve Eq.\eqref{LangevinEq_Rb}-\eqref{LangevinEq_E*} for the steady state number of active ERK molecules ($\bar{E}^*$).

Additionally, at the 32-cell stage  $S_1$ and $S_2$ are related through the empirical relation \cite{williaumeCellGeometrySignal2021}:
\begin{equation}
    S_2=-1.13 S_1+0.91S_{\rm cell}.
    \label{S1S2_relation}
\end{equation} 
Therefore we can compute the number of active ERK molecules  as a function of only the surface of the cell exposed to FGF. 
We assume that all the a-line cells have the same number of receptors $R_T = Q_T = 2000$ and perceive the same extracellular concentrations of FGF and ephrin as in our previous work \cite{williaumeCellGeometrySignal2021, bettoniModelNeuralInduction2023}. We also neglect any diffusion of FGF and ephrin ligands beyond the boundaries of the cell contacts. As a result, the different levels of ERK* predicted by the model in the four a-line cells are a consequence of the different cell surface contacts with FGF and ephrin.

In Fig.~\ref{F1}(c) the experimental measurements of dpERK immunofluorescence (colored dots) obtained in the different a-line cells are compared with the level of ERK* fluorescence predicted with the model (black line) using the parameter values listed in table \ref{tab1}.
The level of ERK* fluorescence (and similarly the noise, shown in gray in the scheme) can be computed from the number of active ERK molecules, as described in section \ref{reproduction_expData}. 
Despite its simplicity, the model successfully reproduces the experimental data obtained in wild type and ephrin-inhibited embryos (see Fig. \ref{F1}(c) and \ref{SupplFig_expData}). 
The experimental data in both figures ~\ref{F1}(c) and \ref{SupplFig_expData}(c) are significantly more spread than the model predictions. As well as intrinsic noise, it is likely that cell to cell variability and experimental measurement noise contribute to the dispersion of the experimental data.

\subsection{Single cell case: the information transmitted between FGF and ERK* depends on the geometry of the cell \label{sec_one-cell}}

\begin{figure*}
    \centering
    \includegraphics[width=1\textwidth]{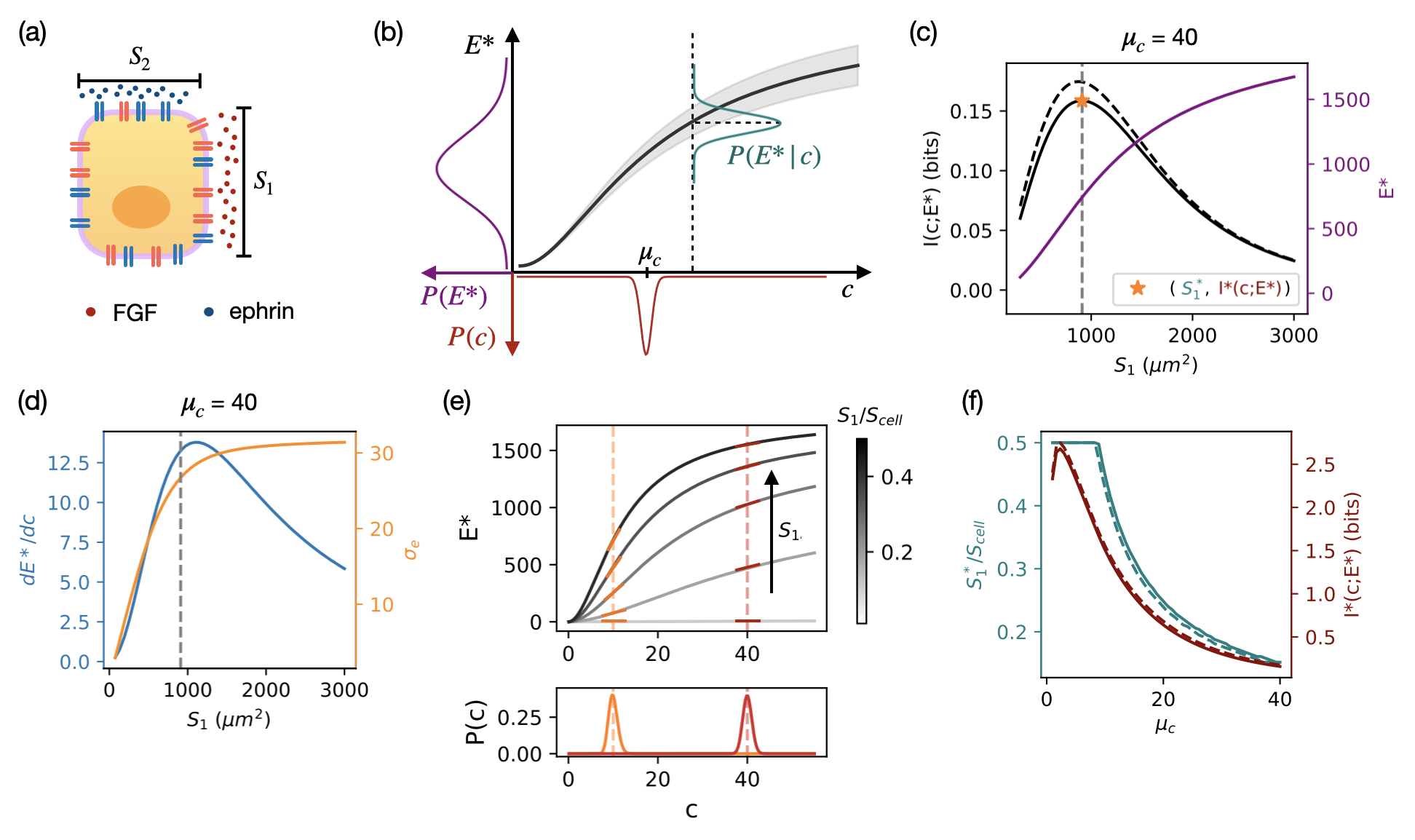}
    \caption{(a) The single cell case. $S_1$ and $S_2$ are the areas of cell surface exposed to FGF and ephrin. The value of $S_1$ is constrained to be at most equal to $S_{\rm cell}/2$, with $S_{\rm cell}$ being the total surface of the cell. The value of $S_2$ is computed from the value of $S_1$ using relation Eq.\eqref{S1S2_relation}. 
(b) Scheme representing the input $P(c)$, conditional $P(E^*\vert c)$ and output $P(E^*)$ distributions. The distribution of the input $P(c)$ is a log-normal distribution centered around $\mu_c$. The conditional distribution $P(E^*\vert c)$ is a Gaussian centered around $\bar{E}^*(c)$, with variance $\sigma_e^2(c)$. $\bar{E}^*(c)$ is the number of active ERK molecules in the cell and is computed with our model as described in section \ref{sec_model}. $\sigma_e(c)$ is the noise in the number of ERK* molecules and is computed using the Langevin approach as described in section \ref{Langevin approach}. 
(c) Information transmitted between the input $c$ and the output $E^*$ as a function of $S_1$ in the presence of ephrin (solid line) and in the absence of ephrin (dashed line). The maximal information (indicated as $I^*(c; E^*)$) is transmitted for $ S_1 \simeq 910$ $\mu m^2$ (indicated with a star). In purple is represented the function $E^*(S_1)$ for $\mu_c = 40 $. 
(d) The plot shows the steepness $\frac{d\bar{E}^*}{dc} \big \vert _{\mu_c}$ (in blue) and the noise $\sigma_e(c = \mu_c)$ (in orange) of the input-output relation $\bar{E}^*(c)$ as a function of $S_1$. The curves are obtained with $\mu_c = 40$. The vertical gray line represents the value of $S_1$ for which the information transmitted $I(c; E^*)$ is maximal.
(e) The main plot shows how the relation between $\bar{E}^*$ and $c$ varies depending on the value of $S_1$.
The value of $S_2$ is computed from the value of $S_1$ through Eq.\eqref{S1S2_relation}. 
The lower plot shows the input distribution $P(c)$ centered at $\mu_c = 10$ (in orange) and at $\mu_c = 40$ (in red). The steepness of the curves close to $c = 10$ and $c= 40$ are shown in orange and red, respectively.
(f) Maximal amount of information $I^*(c; E^*)$ (red curves) as a function of the position $\mu_c$ of the input distribution and the optimal value of $S_1$ (i.e. the value for which information transmission is maximal, green curves) as a function of $\mu_c$,  in the presence (solid lines) and absence (dashed lines) of ephrin. Panels (c), (d) (e) and (f) are obtained assuming: $S_{\rm cell} = 6000 $ $\mu m^2$, $\sigma_c = 1$. }
    \label{fig_2}
\end{figure*}
To assess the influence of the geometry of the system on information transmission, we first analyze the case in which a single cell is exposed to the extracellular ligands FGF and ephrin through the surfaces $S_1$ and $S_2$, as schematized in Fig. \ref{fig_2}(a). In response to the concentrations of FGF and ephrin, a given number of ERK molecules is activated in the cell ($E^*$). The mutual information transmitted between the input ($c$) and the output ($E^*$) can be computed as \cite{shannonMathematicalTheoryCommunication1948}: 
\begin{equation}
\begin{split}
    &I(c;E^*)=- \int  P(E^*) \log_2(P(E^*)) \  dE^* + \\
    & + \int P(c) \ dc \int P(E^*\vert c)  \log_2(P(E^*\vert c)) \ dE^*. 
\end{split}
\end{equation}
The input ($P(c)$), output ($P(E^*)$) and conditional  ($P(E^*\vert c)$) distributions are schematized in Fig. \ref{fig_2}(b). 
Since $c$ cannot take negative values, we assume $P(c)$ to be a log-normal distribution centered around $c = \mu_c$ with variance $\sigma_c^2$:
\begin{equation}
    P(c) = \frac{1}{c\sigma \sqrt{2\pi}} \exp{\left (-\frac{(\ln{(c)}-\mu)^2}{\sigma^2} \right )};
\end{equation}
where: 
\begin{equation}
    \mu = \ln{\left ( \frac{\mu_c^2}{\sqrt{\mu_c^2+\sigma_c^2}} \right )}; \ \ \ \ \ \ 
    \sigma^2 = \ln{\left ( 1 + \frac{\sigma_c^2}{\mu_c^2} \right )}  . 
\end{equation} 
In line with the assumption that there is no gradient of FGF, and that all the a-line cells perceive nearly the same concentration of FGF, we consider the input distribution to be very narrow ($\sigma_c= 1$ compared to the normalized binding constant of the FGF receptors $K_d=k_{d-}/k_{d+} = 60$, which corresponds to 6 nM), see Fig. \ref{fig_2}(b) for a schematic representation. In the absence of more information, we assume the
conditional distribution $P(E^*\vert c)$ to be a Gaussian centered around $E^*= \bar{E}^*(c)$ with variance $\sigma_e^2(c)$: 
\begin{equation}
    P(E^*\vert c)= \mathcal{G}(E^*, \bar{E}^*(c), \sigma_e^2(c));
\end{equation}
where $\bar{E}^*(c)$ is computed solving Eq.\eqref{LangevinEq_Rb}-\eqref{LangevinEq_E*}, at steady state and $\sigma_e^2(c)$ is obtained using the Langevin approach as described in section \ref{Langevin approach}. Finally, the output distribution $P(E^*)$ can be computed directly from the input and conditional distributions as: 
\begin{equation}
    P(E^*)= \int P(E^*\vert c) P(c) \ dc .
\end{equation} 
Therefore, the amount of transmitted information $I(c;E^*)$ is determined once the input $P(c)$ and conditional $P(E^*\vert c)$ distributions are specified.  
Mutual information depends on the geometrical configuration of the cell (on $S_1$ and $S_2$) since both the mean $\bar{E}^*(c)$ and the variance $\sigma_e^2(c)$ of the conditional distribution $P(E^*\vert c)$ are functions of the surface areas $S_1$ and $S_2$. 
At this stage in the embryo, $S_2$ can be computed from the value of $S_1$ through phenomenological Eq.\eqref{S1S2_relation} \cite{williaumeCellGeometrySignal2021}. We can therefore study the dependence on the transmitted information solely as a function of $S_1$. 

We directly compute the mutual information $I(c;E^*)$ for increasing values of $S_1$ (Fig. \ref{fig_2}(c)), for $\mu_c= 40$ which corresponds to 4 nM, and parameters given in table \ref{tab1}. 
The curve obtained in the presence of ephrin (which is the physiological case, shown with a solid line in Fig. \ref{fig_2}(c)) is compared with the curve obtained in the absence of ephrin (shown as a dashed line in Fig. \ref{fig_2}(c)). Both curves have a maximum at intermediate values of $S_1$ ($\simeq 910 $ $\mu m^2$). This may seem counter-intuitive since one could expect the maximum of information transmission to occur at the largest possible value of $S_1$, when the highest number of ERK molecules can be activated. The optimum cell surface area results from assuming a narrow input distribution $P(c)$, which limits the concentration range of the input-output relation in which information is transmitted to the narrow range defined by $c \simeq \mu_c$.
More precisely, the amount of transmitted information $I(c;E^*)$ is determined by the steepness ${d\bar{E}^*}/{dc}(c)$ and the noise $\sigma_e(c)$ of the input-output relation close to $c=\mu_c$. The steepness of the curve determines how many different levels of ERK* can be distinguished by varying the input $c$, hence a higher steepness corresponds to a larger amount of transmitted information.
The noise instead limits the ability to distinguish different levels of the output and therefore higher levels of noise correspond to a lower amount of transmitted information.

When the input distribution is centered around $\mu_c=40$ as in Fig.\ref{fig_2}(c), the steepness of the input-output relation has a peak for intermediate values of $S_1$, while the noise monotonically increases with $S_1$ (see Fig. \ref{fig_2}(d)). 
The optimal trade-off between maximizing the steepness and minimizing the noise is achieved by the surface area $S_1^*$ (vertical gray line in Fig. \ref{fig_2}(d)) at which the transmitted information is maximal.

The dependence of the steepness $d{\bar{E}^*}/{dc}(c)$ (compare orange and red bars in Fig. \ref{fig_2}(e)) and noise  $\sigma_e(c)$ (not shown) on $S_1$ are both a function of the mean input concentration $\mu_c$, which results in different optimal surfaces $S_1^*$ (green curves in Fig. \ref{fig_2}(f)) and maximal amount of information $I^*(c;E^*)$  (red curves in Fig.~\ref{fig_2}(f)) for increasing values of $\mu_c$. For small mean input concentrations, the maximal allowed cell surface area ($S_1 = S_{\rm cell}/2$) is exposed to FGF. Information decreases with  $\mu_c$ since the steepness of the input-output relation decreases (see Fig. \ref{fig_2}(e)) and the noise increases (not shown). 

The presence of ephrin (solid lines in Fig.~\ref{fig_2}(c) and (f)) only slightly reduces the amount of transmitted information (compare solid and dashed lines in Fig.~\ref{fig_2} (c) and (f)) and has very little impact on the value of the optimal surface area $S_1^*$. The reduction of information in the presence of ephrin is most likely due to the fact that ephrin decreases both the number of active ERK molecules and the steepness of the input-output curve $d{\bar{E}^*}/{dc}(c)$.

\subsection{Identical cell case:  symmetry breaking is triggered by the constraint on the total surface \label{symmetric_case}}

\begin{figure*}[ht]
    \centering
    \includegraphics[width=1\textwidth]{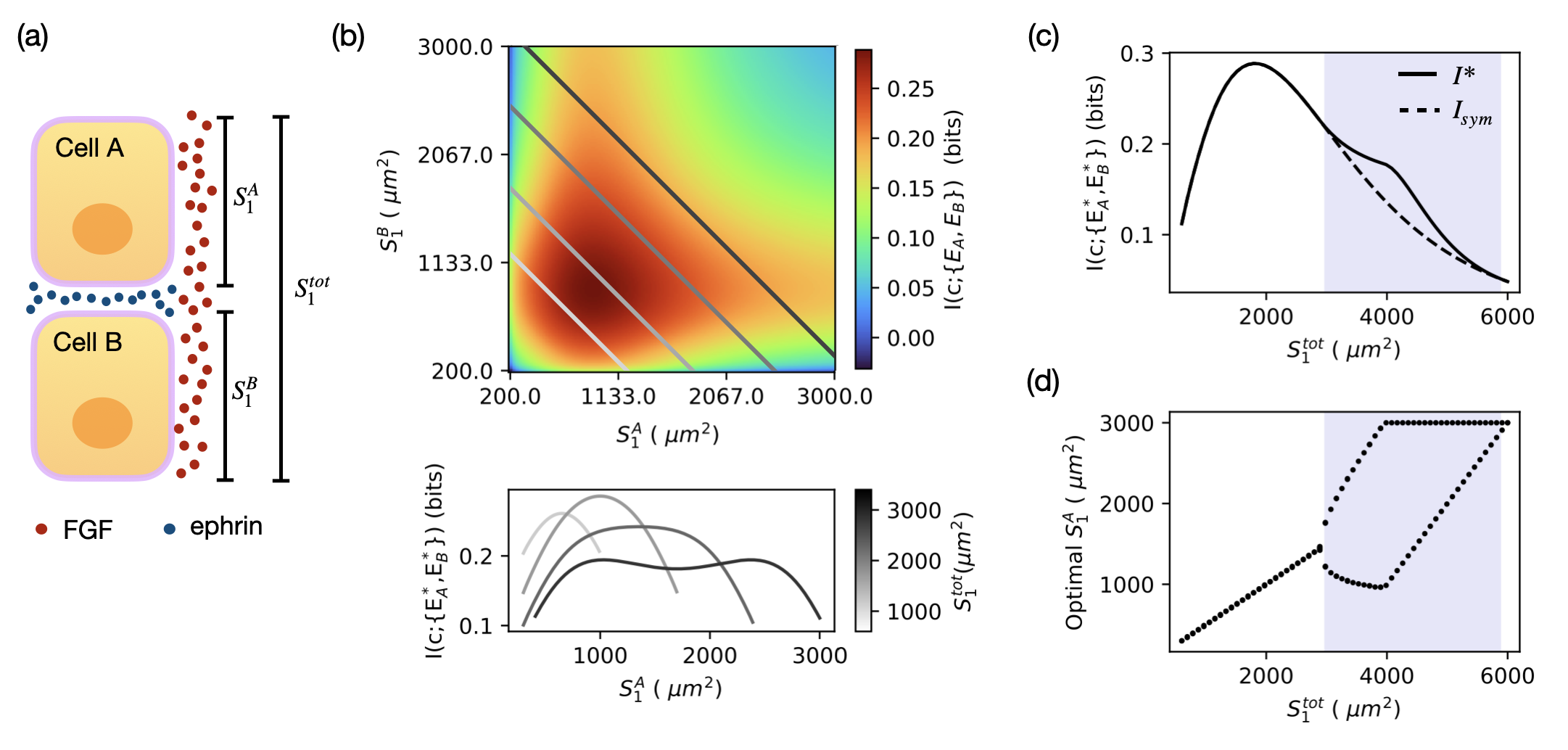}
    \caption{(a) The two-cell case.  $S_1^A$ and $S_1^B$ are the surface areas of the cells exposed to FGF.  $S_2^A$ and $S_2^B$ (not shown in the cartoon) are the surface areas of the cells exposed to ephrin. In each cell, $S_2^{A,B}$ is computed from the corresponding value of $S_1^{A,B}$ using equation Eq.\eqref{S1S2_relation}. The value of $S_1$ in each cell is constrained to be at most equal to $S_{\rm cell}/2$, with $S_{\rm cell}$ being the total surface area of the cell. $S_1^{\rm tot}$ represents the total surface area where FGF is present. The presence of the constraint implies that $S_1^A + S_1^B = S_1^{\rm tot}$. 
(b) The heatmap represents the information transmitted between the FGF input $c$ and the ERK output $\{E^*_A, E^*_B\}$ as a function of the area of cell surfaces exposed to FGF ($S_1^A$ and $S_1^B$). The presence of the constraint restricts the possible values of $S_1^A$ and $S_1^B$ to straight lines defined by $S_1^B = S_1^{\rm tot} -S_1^A$. The straight lines colored with different shades of gray represent the accessible values of $S_1^A$ and $S_1^B$ for the values of $S_1^{tot}$ used in the plot below.
The plot below represents $I(c; \{ E_A^*, E_B^*\})$ as a function of $S_1^A$ for different values of the constraint  $S_1^{\rm tot}$ (see colorbar). 
(c) Maximal information ($I^*$) transmitted between the FGF input  $c$ and the ERK output $\{E^*_A, E^*_B\}$ (solid line, obtained when the two cells expose the optimal surface areas to FGF) and the information transmitted when the two cells have the same surface areas exposed to FGF ($I_{sym}$, dashed line) as a function of $S_1^{\rm tot}$. 
(d) Optimal values of $S_1^{A}$ as a function of  $S_1^{\rm tot}$. The purple background highlights the regions of the plots where it is maximally informative  for the two cells to expose different surface areas to FGF. Panels (b), (c) and (d) are obtained assuming $\mu_c = 40$, $\sigma_c=1$, $S_{\rm cell}^A = S_{\rm cell}^B = \ 6000 \ \mu m^2$.}
    \label{fig_3}
\end{figure*}

We now consider multiple cells exposed to FGF and ephrin. We first assume that all cells are identical, and differ only by the cell surface contacts with FGF and ephrin. 
All the other parameters of the model (including the total surface of the cells) have the same values in all the cells. We add an external constraint $S_1^{\rm tot}$ to take into account the limited area of FGF producing cells in the embryo (Fig. \ref{fig_3}(a)). 

We compute the information transmitted between the input  $c$ and the number of active ERK molecules $E^*$, generalizing the procedure in section~\ref{sec_one-cell} as described in section \ref{meth_info}. We investigate whether there is one (or more) geometrical configuration of the cells that maximizes information transmission and ask how the answer depends on the value of the constraint $S_1^{\rm tot}$. The optimization procedure is described in section~\ref{Optimisation}.

\subsubsection{Two-cell case}

To set up the framework we will first consider a two-cell system (Fig. \ref{fig_3}(a)). The cells are exposed to FGF through the surfaces $S_1^A$ and $S_1^B$ and to ephrin through the surfaces $S_2^A$ and $S_2^B$ (not marked in the cartoon). The number of ERK molecules activated in the two cells in response to the extracellular signals $c$ and $e$ are indicated by $E^*_A$ and $E^*_B$.   
In each cell, the surface area exposed to ephrin ($S_2^{A,B}$) is computed from the corresponding value of $S_1$ using Eq.\eqref{S1S2_relation}.
As a consequence, the information $I(c;\{E^*_A, E^*_B\})$ transmitted between the input $c$ and the outputs $\{E^*_A, E^*_B\}$ depends exclusively on the cell surface area in contact with FGF ($S_1^A$ and $S_1^B$). 
The dependence of $I(c;\{E^*_A, E^*_B\})$ on $S_1^A$ and $S_1^B$ is shown in the heatmap in Fig.~\ref{fig_3}(b) for $\mu_c = 40$. 
At this value of $\mu_c$, the input distribution $P(c)$ can be  approximated with $\delta(c-\mu_c)$, since $\sigma_c << \mu_c$. The two cells therefore are basically independent and $I(c;\{E_A, E_B \}) = I(c; E_A) + I(c; E_B)$ (compare Fig. \ref{fig_3}(b), (c) and (d) with Fig. \ref{SupplFig_independent}(a), (b) and (c)). As a result, the globally optimum surface for the two cells is the same as the one obtained in the single cell case $ S_1^{A*} = S_1^{B*} \simeq 910 $ $\mu m^2$.

The presence of the external constraint $S_1^{tot}$ imposes that $S_1^A + S_1^B = S_1^{\rm tot}$. As a result, it restricts the possible values of $S_1^A$ and $S_1^B$ to straight lines defined by $S_1^B = S_1^{\rm tot}-S_1^A$ (some of these lines are highlighted in the plot with different shades of gray, indicating different values of $S_1^{\rm tot}$). The information $I(c;\{E^*_A, E^*_B\})$ computed along these lines is shown in the lower plot of Fig. \ref{fig_3}(b) as a function of $S_1^A$. 
At low values of $S_1^{\rm tot}$, the transmitted information has a single maximum (light gray curve in Fig.~\ref{fig_3}(b)). In this regime, information transmission is maximized when both cells expose the same area of cell surface to FGF ($S_1^{A*}= S_1^{B*}=S_1^{\rm tot}/2$).
At higher values of $S_1^{\rm tot}$, information has two equivalent maxima, one for small $S_1^A$ (and large $S_1^B$) and another one for large $S_1^A$ (and small $S_1^B$) (dark gray curve in Fig.~\ref{fig_3}(b)). In this case, in order to maximize information transmission the two cells need to break their symmetry and expose different surfaces to FGF. 

This is analyzed in more detail in Fig.\ref{fig_3}(c) and (d). 
The maximal amount of information $I^*(c;\{E^*_A, E^*_B\})$ computed for increasing values of the constraint $S_1^{\rm tot}$ is shown with a solid line in Fig.~\ref{fig_3}(c). The maximal information, transmitted when the cells expose the optimal surface areas to FGF, is compared to the information transmitted when the cells expose the same surface areas to FGF ($I_{sym}$, shown with a dashed line in Fig. ~\ref{fig_3}(c)). 
The purple background highlights the values of $S_1^{\rm tot}$ for which the symmetry breaking favors information transmission. 
The optimal values of $S_1^A$ are shown in Fig. \ref{fig_3}(d) for increasing values of $S_1^{\rm tot}$. For small $S_1^{\rm tot}$, there is a single optimal value for $S_1^A$. 
As $S_1^{\rm tot}$ increases, one cell initially reaches the maximum allowed surface area exposed to FGF ($S_1=S_{\rm cell}/2$) set by the total cell area. At this point, the other cell increases its value of $S_1$ until it also hits the maximum allowed value $S_{\rm cell}/2$ (Fig.~\ref{fig_3}(d)). 

Since the two cells are independent, the symmetry breaking can be simply explained by the properties of the single cell curve (Fig. \ref{fig_2}(c)). More precisely, 
it is the concavity of the single cell curve that determines which geometrical configuration of the cells maximizes information transmission. At low values of $S_1^{tot}$ only the first part of the curve is accessible (see scheme in Fig. \ref{SupplFig_independent}(d)). Since the curve in this region is concave, the information transmitted by two symmetric cells is larger than any other set of $S_1^A$ and $S_1^B$ that satisfies $S_1^A+S_1^B=S_1^{tot}$. As a result, the cells maximize information transmission when $S_1^A = S_1^B= S_1^{tot}/2$. The opposite occurs for large values of $S_1^{tot}$. In this case the accessible region the curve is convex (see scheme in Fig.\ref{SupplFig_independent}(d)) and therefore the two cells maximize information transmission when $S_1^A \ne S_1^B$. 
The transition between these two regimes (i.e., the symmetry breaking) occurs at $S_1^{tot}/2 \simeq 1490 \ \mu m^2$ corresponding to the inflection point of the single cell curve.

We obtain qualitatively the same results in the absence of ephrin (see supplementary Fig. \ref{SupplFig_two_cells}(a)).
Both in the presence and in the absence of ephrin, the amount of transmitted information is quite low (less then $1$ bit), but this is strongly dependent on the position $\mu_c$ of the input distribution, as shown in Fig. \ref{SupplFig_two_cells}(b). 
Not only the amount of information, but also the values of $S_1^{\rm tot}$ for which the symmetry breaking occurs depend on $\mu_c$ (see supplementary Fig. \ref{SupplFig_two_cells}(b)).

The case in which the two cells have different total surfaces ($S_{\rm cell}^{A} = 3000  $  $\mu m^2$, $S_{\rm cell}^{B} = 6000 $ $\mu m^2$) is schematized in Fig. ~\ref{fig:2cells_asymmetrical}(a). 
The dependence of $I(c;\{E^*_A, E^*_B\})$ on $S_1^A$ and $S_1^B$ is shown in the heatmap in Fig.~\ref{fig:2cells_asymmetrical}(b). 
As for the case in which the cells are identical, there is a single optimal value for the two cell surfaces in the absence of the constraint, with the difference that now the optimal values are different for the two cells ($S_1^{A*} \simeq 455 $ $\mu m^2$, $ S_1^{B*} \simeq 910 $ $\mu m^2$). 
The transmitted information $I(c;\{E^*_A, E^*_B\})$ computed for different values of the constraint is shown in the lower plot of Fig. ~\ref{fig:2cells_asymmetrical}(b) as a function of $S_1^A$. 
As can be seen by comparing Fig.\ref{fig_3}(b) and Fig.~\ref{fig:2cells_asymmetrical}(b), the asymmetry introduced in the total cell surfaces distorts the curves. In particular when $S_1^{tot}$ is large (as for the dark gray curve), the two peaks in the curve are no longer equivalent.
As a result, there is one single optimal geometrical configuration of the cells (with $S_1^{A*}<S_1^{B*}$, see Fig.~\ref{fig:2cells_asymmetrical}(c)). Moreover, the optimal surface contacts with FGF ($S_1^{A*}$ and $S_1^{B*}$), are always different, regardless of the value of $S_1^{tot}$ see Fig.~\ref{fig:2cells_asymmetrical}(c)). 

\subsubsection{Four-cell case}
\begin{figure*}[ht]
    \centering
    \includegraphics[width=1\textwidth]{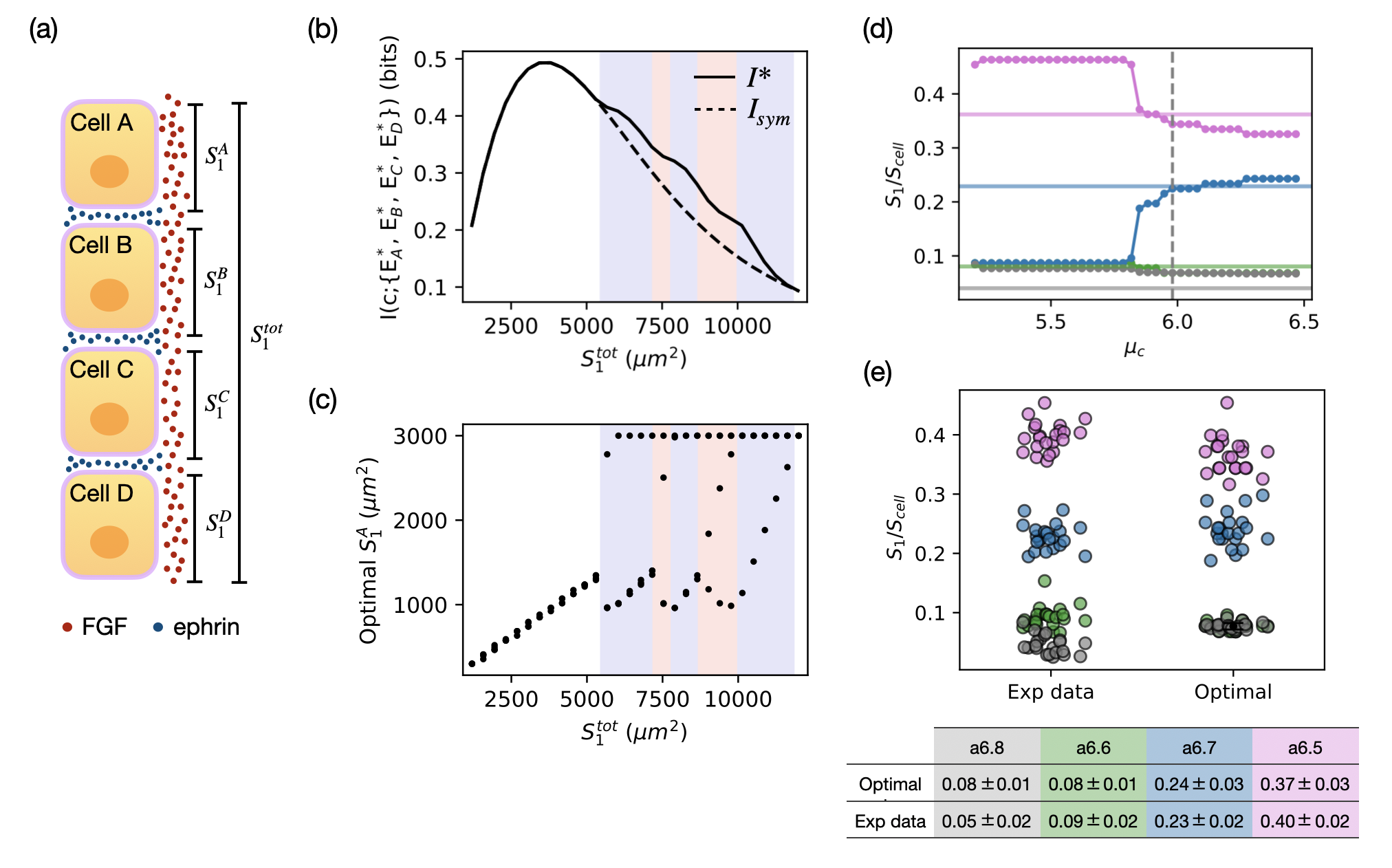}
    \caption{(a) The four-cells case. We consider four cells (labeled A, B, C and D) exposed to FGF and ephrin. In each cell, the surface area exposed to ephrin ($S_2$, not shown in the scheme) is computed from the corresponding value of $S_1$ using equation Eq.\eqref{S1S2_relation}. We assume that the value of $S_1$ in each cell is constrained to be at most equal to $S_{\rm cell}/2$ with $S_{\rm cell}$ being the total surface area of the cell. $S_1^{\rm tot}$ represents the total surface area where FGF is present. The presence of this constraint imposes that $S_1^A + S_1^B + S_1^C + S_1^D = S_1^{\rm tot}$. 
(b) Maximal information ($I^*$) transmitted between $c$ and $\{E^*_A, E^*_B, E^*_C, $ $ E^*_D\}$ (solid line, obtained when the cells expose the optimal surface areas to FGF) and the information transmitted when the four cells expose the same surface area to FGF ($I_{sym}$, dashed line) as a function of $S_1^{\rm tot}$. 
(c) Optimal values of $S_1^{A}$ as a function of  $S_1^{\rm tot}$. In the regions of the plot with purple and red backgrounds there are two or three optimal surface areas, respectively. 
These plots are obtained assuming $\mu_c = 40$, $\sigma_c=1$, $S_{\rm cell}^A = S_{\rm cell}^B = S_{\rm cell}^C = S_{\rm cell}^D = 6000 \ \mu m^2$. (d) Optimal values of the surface areas exposed to FGF ($S_1^{A*}$ in magenta, $S_1^{B*}$ in green, $S_1^{C*}$ in blue and $S_1^{D*}$ in gray) as a function of the position $\mu_c$ of the input distribution. The optimal values are obtained maximizing information transmission in a single embryo using the values of $S_1^{\rm tot}$, $S_{\rm cell}^A$ $S_{\rm cell}^B, S_{\rm cell}^C$ and $S_{\rm cell}^D$ measured experimentally. The horizontal lines represent the values of $S_1$ measured in a6.5 (magenta), a6.6 (green), a6.7 (blue) and a6.8 cells (gray) obtained in the same embryo. The vertical dashed line highlights the value of $\mu_c$ for which the difference between optimal and experimental values is minimal. (e) Comparison between the values of $S_1/S_{\rm cell}$ measured experimentally (on the left) and the closest matching optimal values of $S_1/S_{\rm cell}$ obtained by maximizing information transmission (on the right) in each embryo. The optimal values of $S_1$ are computed using the values of the total cell surface areas measured experimentally in the four a-line cells. The value of $S_1^{\rm tot}$ is also obtained from the experimental data. Each point represents a single cell. The points are colored depending on the cell type (a6.5 in magenta, a6.6 in green, a6.7 in blue and a6.8 in gray). The optimal values are compared with the ones measured experimentally in the table below the plot. For each cell type is computed the mean value of $S_1/S_{cell}$ $\pm$ the standard deviation.}
   \label{fig_4}
\end{figure*}

We now generalize the computation for the four-cell case, which is the condition that resembles what happens in the embryo during neural induction with four anterior animal cells (a6.5, a6.7, a6.6, a6.8) exhibiting different levels of ERK* in response to the extracellular FGF and ephrin signals (Fig. \ref{fig_4}(a)). We consider four cells labeled $A, B, C$ and $D$. The levels of ERK* in the cells are indicated by $E^*_A$, $E^*_B$, $E^*_C$ and $E^*_D$. 
Computing $S_2$ from the value of $S_1$ with Eq.\eqref{S1S2_relation} results in the information $I(c;\{E^*_A, E^*_B, E^*_C, E^*_D\})$ depending only on the cell surface areas in contact with FGF ($S_1^A$, $S_1^B$, $S_1^C$ and $S_1^D$).
Similarly to the two-cell case, the presence of the constraint implies that $S_1^A + S_1^B + S_1^C + S_1^D= S_1^{\rm tot}$. \\
We compute the maximal transmitted information $I^*(c;\{E^*_A, E^*_B, E^*_C, E^*_D\})$  (Fig.~\ref{fig_4}(b)) and the corresponding optimal surface areas ($S_1^{A*}$, $S_1^{B*}$, $S_1^{C*}$ and $S_1^{D*}$,  Fig.~\ref{fig_4}(c)) for increasing values of the constraint $S_1^{\rm tot}$. 
As in the two-cell case, we find that the constraint $S_1^{\rm tot}$ controls the symmetry breaking in the cell surface areas in contact with FGF. More precisely, the value of $S_1^{\rm tot}$ determines whether the optimal geometrical configuration of the cells corresponds to four cells that expose the same surface area $S_1^{A*}= S_1^{B*}= S_1^{C*}= S_1^{D*} =S_1^{\rm tot}/4$ to FGF (regions of Fig.~\ref{fig_4}(b) and (c) with white background) or to four cells that expose different surface areas to FGF (regions of Fig.~\ref{fig_4}(b) and (c) with purple and red backgrounds). The purple regions correspond to values of $S_1^{\rm tot}$ for which there are two optimal values for $S_1^A$. We also find regions (red backgrounds) with three optimal values for $S_1^A$. 
As for the two-cell case, the values of the constraint for which the symmetry breaking occurs, the maximal information transmitted $I^*(c;\{E^*_A, E^*_B, E^*_C, E^*_D\})$ and the optimal surface areas depend on the mean value of the FGF input distribution  $\mu_c$ (not shown).

\subsection{The contact surface areas measured experimentally are compatible with optimal information transmission  \label{sec_four-cells}}

In this section we investigate whether the surface contacts of the a-line cells with FGF-expressing mesendoderm cells 
in the embryo could be optimized 
for information transmission. We perform the same calculations as in the four-cell case described in the previous section, but now we consider that the total surface areas of the four cells ($S_{\rm cell}^{A}$, $S_{\rm cell}^{B}$, $S_{\rm cell}^{C}$, $S_{\rm cell}^{D}$) are equal to the ones measured experimentally in a6.5, a6.6, a6.7 and a6.8 cells. Similarly we consider the value of $S_1^{\rm tot}$ to be equal to the value measured experimentally. 

We compute the optimal surface areas $S_1^{A*}$, $S_1^{B*}$, $S_1^{C*}$  and $S_1^{D*}$ in each embryo and compare them to the values of $S_1$ measured experimentally in the different a-line cells. 
As the optimal surfaces vary depending on the mean value of the input distribution $\mu_c$ (see Fig.\ref{fig_4}(d)), we consider that the value of $\mu_c$ can be slightly different in different embryos. For each embryo we choose the value of $\mu_c$ that minimize the difference between the optimal and experimental data (highlighted with the vertical dashed line in Fig. \ref{fig_4}(d)). The values of $\mu_c$ that match the experimental data ($\mu_c = 5.4 \pm 0.4$) are close to the value of $[\rm FGF]$ used to reproduce the experimental data in Fig\ref{F1}(c) ($c = 5$). 
The computed and measured distributions of $S_1/S_{\rm cell}$ are shown in Fig.~\ref{fig_4}(e). 
The optimal surface areas obtained by maximizing information transmission can be matched surprisingly well to those obtained experimentally. 
The optimal surfaces preserve the order of normalized cell surfaces ($S_1/S_{cell}$) observed experimentally (a6.5 $>$ a6.7 $>$ a6.6 $>$ a6.8) in the vast majority of the embryos (23/25). The two exceptions (not shown in Fig.\ref{fig_4}(e)) are due to the fact that in these embryos cell a6.5 is larger than cell a6.7 ($S_{\rm cell}^{a6.5} > S_{\rm cell}^{a6.7}$). 
The amount of transmitted information is $\simeq 2.4 \pm 0.1$  bits, which is enough to unambiguously specify different levels of ERK* in the four cells, which would require $\log_2(4)=2$ bits.

\section{Discussion} 
Information transmission during ascidian neural induction depends on the geometry of the cells in the embryo, via the surface contacts with FGF-expressing cells. Here we investigated this dependence both in the single (as in section \ref{sec_one-cell}) and multiple (as in sections \ref{symmetric_case} and \ref{sec_four-cells}) cell case. In the multicellular case, the presence of the external constraint $S_1^{\rm tot}$ plays a crucial role in determining the optimal geometrical configurations of the cells. In particular, the constraint determines whether the configuration that maximizes information transmission consists of cells that expose identical (or different) surface areas to extracellular FGF.  

The contact surface areas of the a-line cells in the ascidian embryos at the 32-cell stage could be close to optimal for information transmission
(see Fig. \ref{fig_4}(e)). 
This 
is rather surprising since embryo and molecular configurations are expected to be under multiple selective pressures. Moreover, at the level of the whole developing embryo, the ERK pathway is used to specify a variety of distinct cell types \cite{lemaireUnfoldingChordateDevelopmental2009, biasuz}. 
Any alterations to cell signaling or contacts will likely have an impact on subsequent patterning events. 
For these reasons, we would rather hypothesize that optimization of information transmission would be taking place at the level of the whole embryo patterning. 
Specifically, the quantity that may be optimized is the information transmitted between the morphogens in the egg and the expression levels of the genes needed to specify all the different cell fates.

The observed results may be further refined if we take into account other outputs regulated by the ERK pathway such as $Otx$. Such refinement has been observed for positional information in the context of gap genes in early \textit{Drosophila} development, where each of the gap genes decreases the uncertainty in the readout of the Bcd gradient relative to the positioning of nuclei along the anterior-posterior axis~\cite{petkovaOptimalDecodingCellular2019}.

Optimization of information transmission in neural induction can also be tackled from a different perspective, assuming the input of our system to be the different cell surface contacts with FGF-expressing cells ($S_1$) and not FGF concentration. Experimentally we have access to the distribution of the relative cell surface contacts ($S_1/S_{\rm cell}$). We could have optimized $I(S_1,E^*)$ and compared the optimal distribution of surface contacts with the one measured experimentally. The challenge of this approach consists in the computation of the optimal distribution $P(S_1^A, S_1^B, S_1^C, S_1^D)$. This problem, which is hard to solve analytically because of the presence of multiple inputs, could be addressed numerically by the use of the Blhaut-Arimoto algorithm \cite{blahutComputationChannelCapacity1972}.  
We analyze the case of a single cell in appendix \ref{Opt_IS1}. As shown in Fig.\ref{Fig_appendix2}, the results obtained by optimizing $P(S_1)$ are in qualitative agreement with the ones obtained in section \ref{sec_one-cell}, with the prediction of a single cell surface optimizing information transfer in the presence of a constraint on the area of the cell exposed to FGF.

Further refinement in understanding the interplay between geometry and signaling in ascidian development includes solving the problem of information transmission in realistic 3D geometries,  including  cellular tensions and divisions. Extending the approach of ~\cite{ichbiahEmbryoMechanicsCartography2023} to the 32-cell stage embryo and combining it with the signaling transmission considered here is an interesting, although difficult extension. Another important aspect of embryogenesis, that we did not address in this work, is its reproducibility across embryos. This could be tackled by adopting the approach described in \cite{brucknerInformationContentOptimization2023}.

The importance of the contact surface areas is not only limited to the establishment of the body plan. 
Despite genetic diversity different ascidian species show very similar early development, and in particular the geometric arrangement of the cells across different species is conserved \cite{lemaireEvolutionaryCrossroadsDevelopmental2011, stolfiDivergentMechanismsRegulate2014,dumollardInvariantCleavagePattern2017}. Overall, relying on surfaces of contact for development might be a way of being robust under parameter and genetic changes.

 In summary, optimizing information transmission in neural induction 
 we find that the optimal configuration of the a-line cells is the one in which the cells expose different surfaces to FGF 
 (Fig. \ref{fig_4}(e)). This is the same as what occurs in real embryos in which the a-line cells expose different surface areas to FGF. The surface areas are used by the cells as a knob to tune the level of signal they receive, which is in turn translated into different activation levels of ERK, leading eventually to differential cell fate induction. This approach suggests that ascidian neural induction may be a consequence of the geometrical constraint imposed on the system.

\section{Methods}
The model is implemented in Python. All the codes and the experimental data are available at: \url{https://github.com/rossanabettoni/Information-transmission-NI}. 

\subsection{Stochastic model of ERK activation \label{Langevin approach}} 

\renewcommand{\arraystretch}{1.6}
\begin{table*}	
    \centering
    \begin{tabularx}{0.8\textwidth} {cXc }
       \hline
       \textbf{Parameter}  & \textbf{Definition} & \textbf{Value} \\ \hline 
       $c = \frac{[\rm FGF]}{[ \rm FGF]_0} $  &  Relative extracellular concentration of FGF, $[\rm FGF]_0\simeq 0.1 \ nM$   &  Variable \\ \hline 
       $e = \frac{[ \rm eph]}{[\rm eph]_0}$  &  Relative extracellular concentration of ephrin, $[\rm eph]_0\simeq 0.1 \ nM$  & 5 \\ \hline  
       $E_T$ & Total number of ERK molecules  & 4000 \\ \hline
       $k$  & ERK* de-activation constant &0.2  $s^{-1}$\\   \hline 
       $k_{d-}$ & Unbinding rate constant of FGF to its receptor  & 6 $s^{-1}$  \\ \hline 
       $k_{d+}$ & Binding rate constant of FGF to its receptor multiplied by $[\rm FGF]_0$ &   0.1 $s^{-1}$ \\ \hline 
       $k_{e-}$  & Unbinding rate constant of ephrin to its receptor & 4  $s^{-1}$  \\ \hline 
       $k_{e+}$  & Binding rate constant of ephrin to its receptor multiplied by $[\rm eph]_0$ & 0.1 $s^{-1}$ \\ \hline 
       $K_{rg}$ & Half saturation constant for $Q_b$ &  200 \\ \hline 
       $K_S$ & Half saturation constant for $R_b$ & 200 \\ \hline
       $Q_T$ & Total number of ephrin receptors & 2000 \\ \hline
       $R_T$ & Total number of FGF receptors & 2000 \\ \hline 
       $S_1$ & Surface exposed to FGF & Variable \\ \hline 
       $S_2$ & Surface exposed to ephrin & Variable \\ \hline 
       $S_{\rm cell}$ & Total surface of one cell & 6000 $\mu m^2$ \\ \hline 
       $V_S$ & Maximum rate of ERK activation & 0.2 $s^{-1}$ \\ \hline
       $V_{rg}$ & Maximum rate of ERK deactivation & 0.08 $s^{-1}$ \\ \hline 
    \end{tabularx}
    \caption{Default values of the parameters of the model.}
    \label{tab1}
\end{table*}
\renewcommand{\arraystretch}{1}

To compute the variance in the number of active ERK molecules we use the Langevin approach ~\cite{gillespieChemicalLangevinEquation2000, tkacikInformationFlowOptimization2008}. We linearize equations \eqref{LangevinEq_Rb}-\eqref{LangevinEq_E*} around steady-state and compute the response of the variables $R_b, Q_b$ and $ E^*$ to the noise $ \xi_{R}, \xi_{Q}, \xi_E^*$. 
The Langevin noise terms $\xi_{R}$, $\xi_{Q}$ and $\xi_{E}$ are defined by their statistical properties. They have zero mean: 
\begin{equation*}
    \langle \xi_{R}(t)\rangle = \langle \xi_{Q}(t)\rangle = \langle \xi_{E}(t)\rangle =0
\end{equation*} 
and are uncorrelated in time:
\begin{equation}
\begin{split}
    &\langle \xi_{R}(t)\xi_{R}(t')\rangle \  = A_R \delta(t-t') ; \\
    &\langle \xi_{Q}(t)\xi_{Q}(t')\rangle \  = A_Q \delta(t-t') ;  \\
    &\langle \xi_{E}(t)\xi_{E}(t')\rangle \  = A_E \delta(t-t') ; 
\end{split}
\label{corr_noiseterms}
\end{equation}
where the noise amplitudes $A_R$, $A_Q$, $A_{E}$ are given in Eqn~\ref{AE}, ~\ref{An} and ~\ref{Am}.

The evolution equations for small departures from steady-state are: 
\begin{equation}
    \dot{\delta R}_b =  -(k_{d\scriptscriptstyle+}c + k_{d\scriptscriptstyle-} ) \delta R_b + \xi _{R} ; 
    \label{eq_deltan_i}
\end{equation}
\begin{equation}
 \dot{\delta Q}_b=  -(k_{e\scriptscriptstyle+}e + k_{e\scriptscriptstyle-} ) \delta Q_b + \xi _{Q} ; 
 \label{eq_deltam_j}
\end{equation}
\begin{equation}
        \dot{\delta E^*} =  \Gamma_R \delta R_b -  \Gamma_Q \delta Q_b - \tau_E^{-1} \delta E^* + \xi_{E} ; 
        \label{eq_deltaE*}
    \end{equation}
where: 
\begin{equation*}
         \Gamma_R = 2\ V_S\frac{\bar{R}_b}{(\bar{R}_b^2 + K_S^2)^2} (E_T-\bar{E}^*) ; 
\end{equation*}
\begin{equation*}
         \Gamma_Q = V_{rg}\frac{K_{rg}}{(\bar{Q}_b + K_{rg})^2} \bar{E}^* ; 
\end{equation*}
\begin{equation*}
\tau_E^{-1} =  V_s \frac{\bar{R}_b^2}{\bar{R}_b^2 + K_s^2} + V_{rg}\frac{\bar{Q}_b}{\bar{Q}_b + K_{rg}} + k. 
\end{equation*}
The steady-state values for the number of bound FGF and ephrin receptors are given by: 
\begin{equation*}
    \bar{R}_b = R \frac{c}{c+K_d}; \ \ \ \ \bar{Q}_b = Q\frac{e}{e+K_e};
\end{equation*}
where $K_d = k_{d\scriptscriptstyle-}/k_{d\scriptscriptstyle+}$ and $K_e = k_{e\scriptscriptstyle-}/k_{e\scriptscriptstyle+}$. 

Equations \eqref{eq_deltan_i}-\eqref{eq_deltaE*} can be solved introducing the Fourier transforms: 
\begin{equation*}    
\delta R_b(t) = \int \frac{d\omega}{2\pi} e^{-i\omega t} \delta \tilde{R}_b ( \omega ) ; 
\end{equation*}
\begin{equation*}       
\delta Q_b( t) = \int \frac{d\omega}{2\pi} e^{-i\omega t} \delta \tilde{Q}_b ( \omega ) ; 
\end{equation*}
\begin{equation*}
\delta E^*( t) = \int \frac{d\omega}{2\pi} e^{-i\omega t} \delta \tilde{E}^* ( \omega ) . 
\end{equation*}
Similarly, the Langevin noise terms can be written with the Fourier representation: 
\begin{equation*}
    \xi (t) = \int \frac{d\omega}{2\pi} \tilde{\xi} (\omega) . 
\end{equation*}
Solving equation \eqref{eq_deltan_i} for $\delta \tilde{R}_b$ we obtain : 
\begin{equation}
    \delta \tilde{R}_b (\omega) = \frac{\tilde{\xi} _{R} }{-i\omega + \tau_c^{-1}} ; 
    \label{Eq_deltatilde_Rb}
\end{equation}
where $\tau_c^{-1} = (k_{d\scriptscriptstyle+}c + k_{d\scriptscriptstyle-} )$. 
Solving equation \eqref{eq_deltam_j} for $\delta \tilde{Q}_b$ we obtain: 
\begin{equation}
    \delta \tilde{Q}_b (\omega) = \frac{\tilde{\xi} _{Q} }{-i\omega + \tau_e^{-1}} ; 
    \label{Eq_deltatilde_Qb}
\end{equation}
where $\tau_e^{-1} = (k_{e\scriptscriptstyle+}e + k_{e\scriptscriptstyle-} )$. 
Finally, solving equation \eqref{eq_deltaE*} we obtain: 
\begin{equation}
    \delta \tilde{E}^*(\omega) = \frac{\Gamma_R \delta \tilde{R}_b}{-i\omega + \tau_E^{-1}} - \frac{\Gamma_Q \delta \tilde{Q}_b}{-i\omega + \tau_E^{-1}} + \frac{ \tilde{\xi}_{E}}{-i\omega + \tau_E^{-1}} . 
\label{Eq_deltatildeE*}
\end{equation} 
The noise power spectral density of the protein copy number is: 
\begin{equation}
    S_E(\omega) = \langle\delta \tilde{E}^*(\omega) \delta \tilde{E}^{*\dagger}(\omega) \rangle 
    \label{S_E_def} ; 
\end{equation}
where the $\dagger$ indicates the complex conjugate. 
\begin{equation}
\begin{split}
    S_E&(\omega) = \frac{\Gamma_R^2  \langle\delta \tilde{R}_b(\omega) \delta \tilde{R}_b^{\dagger}(\omega) \rangle }{\omega^2 + \tau_E^{-2}} + \\
    + & \frac{\Gamma_Q^2  \langle\delta \tilde{Q}_b(\omega) \delta \tilde{Q}_b^{\dagger}(\omega) \rangle}{\omega^2 + \tau_E^{-2}} + \frac{ \langle\tilde{\xi}_{E}(\omega) \delta \tilde{\xi}_{E}^{\dagger}(\omega) \rangle}{\omega^2 + \tau_E^{-2}} . 
    \label{S_E}
\end{split}
\end{equation}
$  \langle\delta \tilde{R}_b(\omega) \delta \tilde{R}_b^{\dagger}(\omega) \rangle $ can be computed from Eq.\eqref{Eq_deltatilde_Rb} as: 
\begin{equation}
     \langle\delta \tilde{R}_b(\omega) \delta \tilde{R}_b^{\dagger}(\omega) \rangle  
    =   \frac{\langle\tilde{\xi}_{R}(\omega) \tilde{\xi}_{R}^{\dagger}(\omega) \rangle}{\omega^2 + \tau_c^{-2} } . 
\end{equation}
Similarly: 
\begin{equation}
     \langle\delta \tilde{Q}_b(\omega) \delta \tilde{Q}_b^{\dagger}(\omega) \rangle  = \frac{ \langle\tilde{\xi}_{Q}(\omega) \tilde{\xi}_{Q}^{\dagger}(\omega) \rangle}{\omega^2 + \tau_e^{-2} } . 
\end{equation}
From the definition of the Langevin forces (Eq. \eqref{corr_noiseterms}), it follows: 
\begin{equation*}
\begin{split}
    &\langle\tilde{\xi}_{R}(\omega) \delta \tilde{\xi}_{R}^{\dagger}(\omega) \rangle = A_{R}  ;\\
    & \langle\tilde{\xi}_{Q}(\omega) \delta \tilde{\xi}_{Q}^{\dagger}(\omega) \rangle = A_{Q} ; \\
    &\langle\tilde{\xi}_{E}(\omega) \delta \tilde{\xi}_{E}^{\dagger}(\omega) \rangle = A_{E} .
\end{split}
\end{equation*}
Therefore Eq.\eqref{S_E} becomes:  
\begin{equation}
    \begin{split}
        & S_E(\omega) = \frac{ \ \Gamma_R^2 A_R }{(\omega^2 + \tau_E^{-2})(\omega^2 + \tau_c^{-2})} + \\
    & + \frac{\Gamma_Q^2  A_Q}{(\omega^2 + \tau_E^{-2})(\omega^2  + \tau_e^{-2})} + \frac{ A_E}{\omega^2 + \tau_E^{-2}} . 
    \end{split}
\end{equation}
Each Langevin term in the expression above represents a noise source: the first two terms represent the noise in the binding and unbinding of the FGF and ephrin receptors, while the last term represents the noise in the activation of ERK. \\ 
The amplitude $A_E$ can be computed following the procedure described in \cite{gillespieChemicalLangevinEquation2000} as: 
\begin{equation}
    A_E = 2V_S \frac{ \bar{R}_b^2}{\bar{R}_b^2 + K_s^2} (E_T-\bar{E}^*) .  \label{AE} 
\end{equation}
The amplitude $A_R$ can be computed considering that each receptor binding site can be either occupied or empty. The probability that a collection of $R$ receptor binding sites are occupied follows the binomial distribution, therefore the variance of $\delta R_b$ must be given by: 
\begin{equation}
    \sigma_{R_b}^2 = \langle (\delta R_b)^2\rangle = R \ \bar{n}(1-\bar{n}) . 
\end{equation}
where $\bar{n} = \frac{\bar{R}_b}{R}$ and $m=\frac{\bar{Q}_b}{Q}$ are the fractional occupancy of FGF and ephrin receptors, respectively. 
At the same time, the variance can be also computed as: 
\begin{equation}
    \langle (\delta R_b)^2\rangle \ = \int \frac{d\omega}{2\pi} S_{R_b}(\omega) ; 
\end{equation}
where $S_{R_b}(\omega)$ is the power spectral density of the number of bound receptors, that can be computed as $ S_{R_b}(\omega) = \langle \delta \tilde{R}_b(\omega) \delta \tilde{R}_b^\dagger(\omega)\rangle$. Comparing the two expressions for the variance one obtains: 
\begin{equation}
     A_R = \frac{2}{\tau_c} R \ \bar{n} (1-\bar{n}) . \label{An} 
\end{equation}
Similarly, $A_Q$ can be computed as: 
\begin{equation}
    A_Q = \frac{2}{\tau_e} Q \ \bar{m} (1-\bar{m}). \label{Am} 
\end{equation}
The total variance of the protein copy number is given by an integral over the spectrum: 
\begin{equation}
   \langle (\delta E^*)^2 \rangle =  \sigma^2_E = \int \frac{d\omega}{2\pi} S_E(\omega) . 
\end{equation}
The noise was computed numerically as a function of $c$ and/or $S_1$. 

\subsection{Reproduction of the experimental results \label{reproduction_expData}}
Experimentally, ERK activation levels are quantified by immunofluorescence (IF) signals \cite{williaumeCellGeometrySignal2021}. Thus, to compare the simulation results with experiments we considered that the level of ERK fluorescence ($E_f$) is a linear function of the number of active ERK molecules ($E^*$): 
\begin{equation*}
E_f = A \cdot \frac{E^*}{E_T} + B ; 
\end{equation*}
where A is the maximal ERK IF signal and B is the background value of the IF signal. 

The propagation of the intrinsic noise $\sigma_E$ to the level of ERK fluorescence is computed as: 
\begin{equation}
    \Delta E_f = \frac{\partial E_f}{\partial E^*} \sigma_E ; 
\end{equation}
where $\sigma_E$ is computed through the Langevin approach as described in section \ref{Langevin approach}. 

Parameter values were obtained by manual fitting to get best agreement with experimental observations. The default values of the parameters are listed in table \ref{tab1}. 
\newline

\subsection{Computation of mutual information in the multiple cell case. \label{meth_info}}

We consider $N$ different cells in contact with FGF and ephrin through the surfaces $S_1^i$ and $S_2^i$ ($i = 1, .... , N$).  
In response to the extracellular FGF concentration (indicated as $c$) different numbers of ERK molecules will be activated in the different cells. Let the number of ERK* molecules in the $i$-th cell be $E^*_i$. 
It is possible to define the mutual information $I(c;\{ E^*_1, .., E^*_N\})$ transmitted between the input $c$ and the output $\{E^*_1, .. , E^*_N\}$ as: 
\begin{widetext}
\begin{equation}
\begin{split}
    I(c;\{ E^*_1, .., E^*_N\}) & =- \int  P(\{E^*_1, ..., E^*_N\}) \log_2(P(\{E^*_1, ..., E^*_N\}))\ \ dE^*_1 ... dE^*_N + \\
    &+ \int P(c) \ dc \int  P(\{E^*_1, ..., E^*_N\}\vert c)  \log_2(P(\{E^*_1, ..., E^*_N\}\vert c)) \ dE^*_1 ... dE^*_N; 
\end{split} \label{info}
\end{equation} 
\end{widetext}
where $P(c)$ is the distribution of the input, $P(\{E^*_1, ..., E^*_N\})$ is the output distribution and $P(\{E^*_1, ..., $ $E^*_N\}\vert c))$ is the conditional distribution of the output given the value of the input. 
We assume the input distribution $P(c)$ to be a log-normal distribution centered around $\mu_c$ with variance $\sigma_c^2$ as described in section \ref{sec_one-cell}.
Since the cells are not interacting the conditional distribution $P(\{E^*_1, ..., E^*_N\} \vert c)$ can be written as: 
\begin{equation}
    P(\{E^*_1, ..., E^*_N\} \vert c)= \prod_{i=1}^{N}P(E^*_i\vert c).
\end{equation}
In the absence of more information, we assume that the conditional distributions $P(E^*_i\vert c)$ are Gaussian: 
\begin{equation}
    P(E_i^*\vert c)=\frac{1}{\sqrt{2\pi(\sigma_e^i(c))^2}} e^{-\frac{1}{2}\frac{(E_i^*-\bar{E}_i^*(c))^2}{(\sigma_e^i(c))^2}};
\end{equation}
where $\bar{E}^*_i(c)$ is the number of ERK molecules activated in cell $i$, which is computed solving equations Eq.\eqref{LangevinEq_Rb}-\eqref{LangevinEq_E*} at steady-state. $\sigma_e^i(c)$ is the noise in the number of active ERK molecules in the $i$-th cell  which is computed using the Langevin approach as described in section \ref{Langevin approach}. 

Finally, the output distribution can be computed directly from the input and the conditional distribution as: 
\begin{equation}
    P(\{ E^*_1, ..., E^*_N\})=\int \prod_{i=1}^{N}P(E^*_i\vert c) P(c) \ dc. 
\end{equation}
All the integrals were evaluated numerically using the composite trapezoidal rule, implemented with the function numpy.trapz(). 

\subsection{Optimization of information transmission \label{Optimisation}}
In the multiple cells case, we considered $N$ cells exposed to FGF and ephrin through the surfaces $S_1^i$ and $S_2^i$ ($i = 1, .... , N$). In each cell, the surface exposed to ephrin ($S_2^i$) is computed from the corresponding value of $S_1^i$ using equation Eq.\eqref{S1S2_relation}. We assume that value of $S_1$ in each cell is constrained to be at most equal to $S_{\rm cell}/2$ (with $S_{\rm cell}$ being the total surface of the cell). The limiting case in which $S_1^i = S_{\rm cell}/2$ would correspond to the unrealistic condition in which the cell is completely flat.
To mimic the presence of the vegetal cells in the embryo, that produce FGF and define a fixed surface in which FGF is present,
we consider the presence of an external constraint $S_1^{\rm tot}$, given by the following equation: 
\begin{equation}
    \sum_{i=1}^N S_1^i = S_1^{\rm tot} . 
\end{equation}
We computed the optimal surfaces of the cells $S_1^{*1}, ... S_1^{N*}$ that maximize information transmission between the input $c$ and the outputs $\{ E^*_1, ..., E^*_N\}$ and the corresponding maximal amount of information $I^*(c;\{ E^*_1, .., E^*_N\})$. 
The constrained optimization was carried out in Python either by brute-force (in the two-cells case) or by using the Constrained Optimization BY Linear Approximation (COBYLA) method (in the three and four-cells cases) which is an iterative method for derivative-free constrained optimization. 

\subsection{Experimental methods}
All the experimental results are taken from our previous work \cite{williaumeCellGeometrySignal2021}.

\begin{acknowledgments}
We thank very warmly Clare Hudson and Hitoyoshi Yasuo for their thoughtful comments about the manuscript and for answering many of our questions about ascidians. 
RB is supported by a FRIA fellowship. GD is Research Director at the Belgian "Fonds National pour la Recherche Scientifique" (FRS-FNRS) and acknowledges financial support from the ARC project "Noise sensitivity of gene regulatory networks underlying cell fate specification" financed by the Universit\'e libre de Bruxelles (ULB). 
AMW is supported by ANR Distant and CZI Biohub Theory Initiative. The funders had no role in study design, data collection and analysis, decision to publish, or preparation of the manuscript.
The resources and services used in this work were provided by the VSC (Flemish Supercomputer Center), funded by the Research Foundation - Flanders (FWO) and the Flemish Government.
\end{acknowledgments}

\bibliography{biblio}

\newpage

\appendix*

\section{ Optimizing information between $S_1$ and $E^*$ \label{Opt_IS1}}

In this work we maximized the information transmitted between the extracellular [FGF] and the number of active ERK* molecules to find the optimal geometrical configuration of the cell (i.e., the optimal value of $S_1$). The same problem of optimizing information transmission can be tackled considering the surface $S_1$ as the input of the signaling cascade. In this case we can compute the information $I(S_1, E^*)$, and find the optimal distribution of the input $P^*(S_1)$. 
We will describe below the computation in the case of a single cell. To perform the optimization we follow the procedure described in \cite{tkacikInformationTransmissionGenetic2011, tkacikInformationFlowOptimization2008, tkacikOptimizingInformationFlow2009}. When considering multiple cells the analytical approach becomes challenging and numerical techniques must be used. We leave this more complicated problem for future studies. 

The information transmitted between $S_1$ and $E^*$ can be written as: 
    \begin{equation}
\begin{split}
    &I(S_1;E^*)=- \int  P(E^*) \log_2(P(E^*))   dE^* + \\
    & + \int P(S_1)  dS_1 \int P(E^*\vert S_1)  \log_2(P(E^*\vert S_1))  dE^*. 
\end{split}
\end{equation}
We assume that the conditional distribution $P(E^*\vert S_1)$ to be Gaussian: 
\begin{equation}
    \mathcal{G}(E^*, \bar{E}^*(S_1), \sigma_e^2(S_1));
\end{equation}
where $\bar{E}^*(S_1) $ and $ \sigma_e^2(S_1)$ are computed as described in section \ref{sec_model} and \ref{Langevin approach}. 

The integral $\int P(E^*\vert S_1)  \log_2(P(E^*\vert S_1)) \ dE^*$ is simply (minus) the entropy of the Gaussian distribution, which can be computed as $\log_2(\sqrt{2\pi e }\sigma_e(S_1) )$. 

The output distribution can be computed as: 
\begin{equation}
    P(E^*) = \int P(E^*\vert S_1) P(S_1) \ dS_1.
\end{equation}
In the small noise limit, we can approximate $P(E^*\vert S_1)$ with a delta function: 
\begin{equation}
    P(E^*\vert S_1) \simeq \delta(E^* - \bar{E}^*(S_1) ) . 
\end{equation}
As a result, we can rewrite the output distribution $P(E^*)$ in terms of the input distribution $P(S_1)$ as: 
\begin{equation}
    P(E^*) \simeq P(S_1) \left \vert  \frac{d\bar{E}^*}{dS_1}\right \vert . 
\end{equation}
Information therefore becomes: 
\begin{equation}
    I(S_1; E^*)  = \int P(S_1) \log_2\left ( \frac{\vert d\bar{E}^*/dS_1\vert}{   \sqrt{2\pi e } \sigma_e P(S_1)} \right ) dS_1 \label{Info_S1}.
\end{equation}

\subsection{In the absence of external constraints}

\begin{figure}[!ht]
    \centering
    \includegraphics[width=0.9\linewidth]{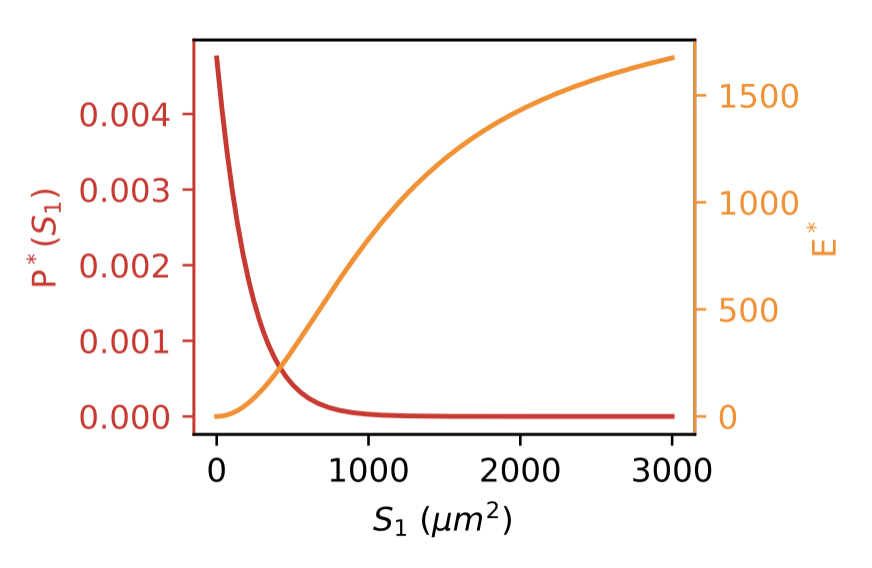}
    \caption{Optimal input distribution $P^*(S_1)$ (in red). In orange is shown the input-output relation $E^*(S_1)$. The plot is obtained assuming $S_{cell} = 6000$ $\mu m^2$, $c = 40$. The value of $S_1$ is constrained to be at most equal to $S_{\rm cell}/2$. }
    \label{Fig_appendix1}
\end{figure}

To maximize information we write the Lagrangian: 
\begin{equation}
\begin{split}
 \mathcal{L}[P(S_1)] = I&(S_1; E^*)  - \Lambda \left (\int P(S_1) dS_1 -1 \right) ; 
\end{split}
\end{equation}
where $\Lambda$ is a Lagrange multiplier that will enforce the normalization of the input distribution $P(S_1)$. 
The optimal input distribution $P^*(S_1)$ must satisfy:
\begin{equation}
    \frac{\partial \mathcal{L}[P(S_1)]}{\partial P(S_1)} = 0.
\end{equation}
As a result, the optimal distribution is given by: 
\begin{equation}
    P^*(S_1) = \frac{1}{Z} \bigg \vert\frac{dE^* }{dS_1} \bigg \vert \frac{1}{\sigma_e(S_1)},
\end{equation}
where the normalization constant $Z$ is: 
\begin{equation*}
    Z = \int \bigg \vert\frac{dE^* }{dS_1} \bigg \vert \frac{1}{\sigma_e(S_1)} dS_1 . 
\end{equation*}
The maximal mount of information is obtained inserting the expression for $P^*(S_1)$ in Eq.\eqref{Info_S1}: 
\begin{equation}
    I^*(S_1; E^*) = \log_2(Z).
\end{equation}
The optimal input distribution (shown in red) is compared with the input-output curve $E^*(S_1)$ (shown in orange) in Fig. \ref{Fig_appendix1}.
\\

\subsection{In the presence of the constraint $\langle S_1 \rangle = S_1^{tot}$}
\begin{figure}[!ht]
    \centering
    \includegraphics[width=0.9\linewidth]{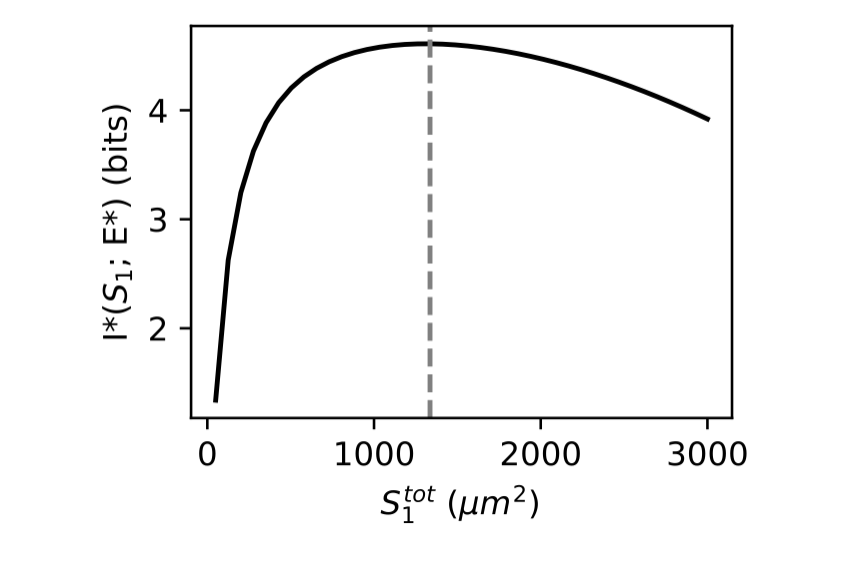}
    \caption{Maximal information $I^*(S_1; E^*)$ transmitted for increasing values of $S_1^{tot}$. The vertical gray line highlights the position of the maximum (which occurs at $\langle  S_1\rangle \simeq 1340 $ $\mu m^2$). The plot is obtained assuming $S_{cell} = 6000$ $\mu m^2$, $c = 40$. The value of $\langle S_1\rangle = S_1^{\rm tot}$ is constrained to be at most equal to $S_{\rm cell}/2$. }
    \label{Fig_appendix2}
\end{figure}

We now consider the presence of an external constraint that fixes the average area of the cell surface in contact with FGF : $\langle S_1\rangle = S_1^{tot}$. 
In the presence of the constraint the Lagrangian becomes: 
\begin{equation}
\begin{split}
 \mathcal{L}[P(S_1)] = I&(S_1; E^*)  - \Lambda \left (\int P(S_1) dS_1 -1 \right) + \\
   &  - \Gamma  \left (\int S_1 P(S_1) dS_1  -S_1^{tot}\right) ;
\end{split}
\end{equation}
where $\Gamma$ is the Lagrange multiplier that fixes the average surface of the cell $\langle S_1\rangle = S_1^{tot}$.
The optimal input distribution in this case becomes: 
\begin{equation}
    P^*(S_1) = \frac{1}{Z} \bigg \vert\frac{dE^* }{dS_1} \bigg \vert \frac{1}{\sigma_e(S_1)} e^{-\Gamma S_1},
\end{equation}
with: 
\begin{equation*}
    Z = \int \bigg \vert\frac{dE^* }{dS_1} \bigg \vert \frac{1}{\sigma_e(S_1)} e^{-\Gamma S_1} dS_1.
\end{equation*}
The maximal information is: 
\begin{equation}
    I^*(S_1; E^*) = \log_2(Z) + \Gamma S_1^{tot} .
\end{equation}
The information transmitted as a function of $S_1^{tot}$ is shown in Fig. \ref{Fig_appendix2}. The curve has a maximum for intermediate values of $\langle S_1\rangle \simeq 1340 $ $\mu m^2$, qualitatively in agreement with the results obtained for the single cell case in section \ref{sec_one-cell}. 

\newpage
\
\newpage

\renewcommand{\thefigure}{S\arabic{figure}}
\setcounter{figure}{0}
\onecolumngrid
\section*{SUPPLEMENTARY MATERIAL}
\begin{figure}[ht!]
    \centering
    \includegraphics[width=1\textwidth]{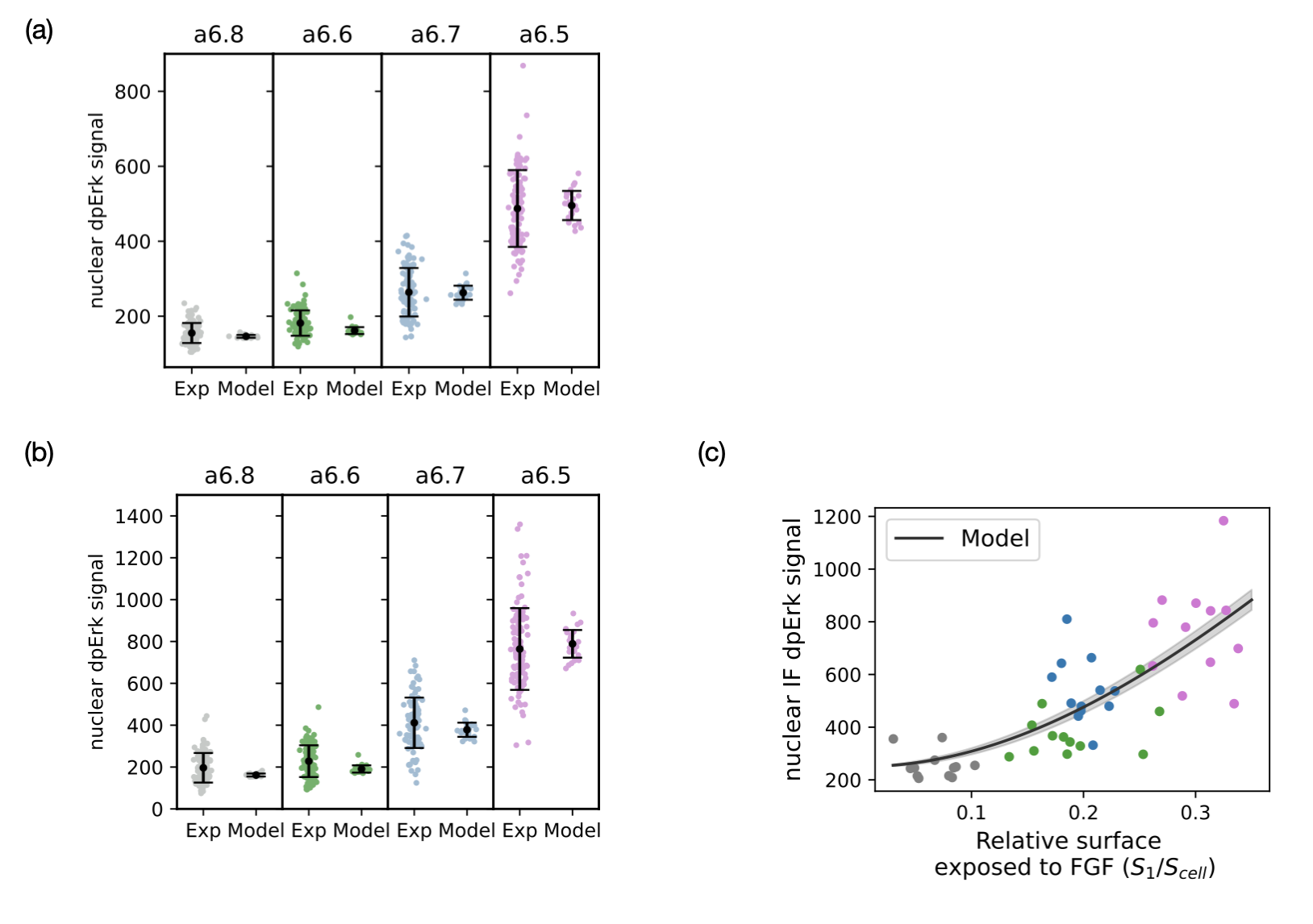}
    \caption{(a) Nuclear dpERK immunofluorescence (IF) signals in the a6.5, a6.6, a6.7 and a6.8 cell types as measured in IF experiments in wild type embryos (left) and computed with the model (right). Each point represents a single cell and modeling results are computed using the measured values of S1. Means and standard deviations are shown in black. The parameter values used to reproduce the experimental data are listed in table \ref{tab1}. $A = 5000$, $B=140$.
    (b) Nuclear dpERK immunofluorescence (IF) signals in the a6.5, a6.6, a6.7 and a6.8 cell types as measured in IF experiments in embryos in which the ephrin pathway was inhibited (left) and computed with the model (right). Each point represents a single cell and modeling results are computed using the measured values of S1. Means and standard deviations are shown in black. 
    The parameter values used to reproduce the experimental data are listed in table \ref{tab1}. $A = 8000$, $B=150$.
    To mimic the absence of ephrin, we imposed $e=10^{-5}$.
    (c) Experimental data in the absence of ephrin. The plot shows the nuclear dpERK immunofluorescence (IF) signals measured experimentally in individual a-line cells as a function of the relative area of cell surface contact with FGF-expressing cells ($S_1/S_{cell}$). Experimental data obtained in the different cell types are shown with dots of different colors: a6.8 cells are shown in gray, a6.6 cells in green, a6.7 cells in blue, and a6.5 cells in magenta. The experimental data obtained in embryos in which the ephrin pathway is inhibited are compared with the model predictions (shown with a black line). The shaded region represents the noise in the level of ERK* fluorescence predicted by the model using the Langevin approach. The model predictions are obtained as described in section \ref{sec_model}, \ref{Langevin approach} and \ref{reproduction_expData}.
    The parameter values used to reproduce the experimental data are listed in table \ref{tab1}. $A = 10000$, $B=250$.
    To mimic the absence of ephrin, we imposed $e=10^{-5}$. 
    }
    \label{SupplFig_expData}
\end{figure}
\vfill

\begin{figure}[ht]
    \centering
    \includegraphics[width=0.9\textwidth]{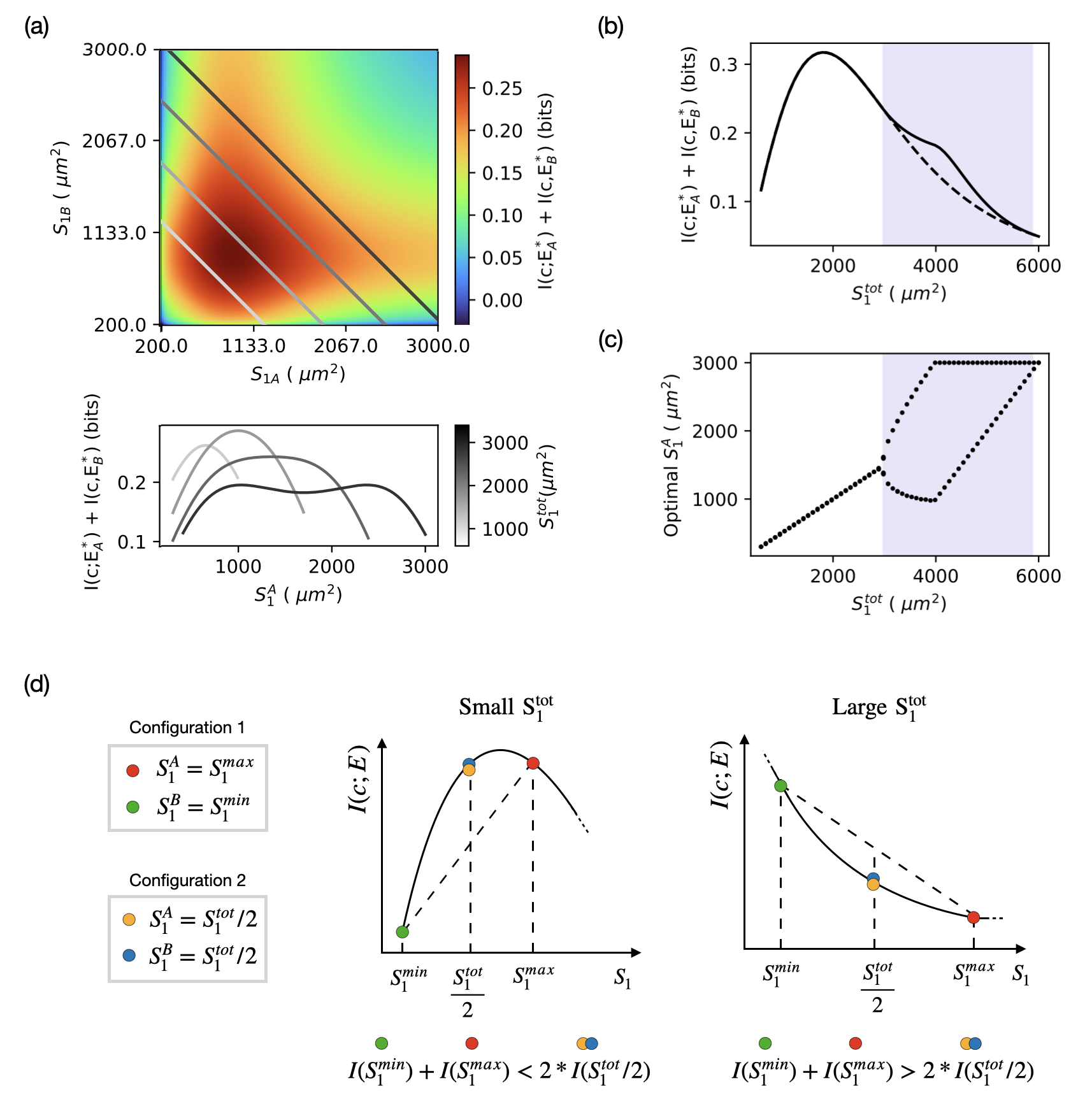}
    \caption{(a) Independent cell case. The heat map represents the information transmitted assuming the two cells are independent  $I = I(c; E_A) + I(c; E_B)$ as a function of the area of cell surfaces exposed to FGF ($S_1^A$ and $S_1^B$). The presence of the constraint restricts the possible values of $S_1^A$ and $S_1^B$ to straight lines defined by $S_1^B = S_1^{\rm tot} -S_1^A$. The straight lines colored with different shades of gray represent the accessible values of $S_1^A$ and $S_1^B$ for the values of $S_1^{tot}$ used in the plot below.
The plot below represents $I= I(c; E_A) + I(c; E_B)$ as a function of $S_1^A$ for different values of the constraint  $S_1^{\rm tot}$ (see colorbar). 
(b) Maximal information (solid line, obtained when the two cells expose the optimal surface areas to FGF) and the information transmitted when the two cells have the same surface areas exposed to FGF (dashed line) as a function of $S_1^{\rm tot}$. 
(c) Optimal values of $S_1^{A}$ as a function of  $S_1^{\rm tot}$. The purple background highlights the regions of the plots where it is maximally informative  for the two cells to expose different surface areas to FGF. Panels (a), (b) and (c) are obtained assuming $\mu_c = 40$, $\sigma_c=1$, $S_{\rm cell}^A = S_{\rm cell}^B = \ 6000 \ \mu m^2$. 
(d) Scheme illustrating why the optimal geometrical configuration of the cells vary with $S_1^{\rm tot}$. We compare two geometrical configurations of the cells:  configuration 1 in which $S_1^A = S_1^{max}$ and $S_1^B = S_1^{min}$, and configuration 2 in which $S_1^A = S_1^B = S_1^{\rm tot}/2$. $S_1^{max}$ and $S_1^{min}$ are the maximal and minimal possible values for $S_1$ (fixed by the values of $S_{cell}$ and $S_1^{\rm tot}$).  At low values of $S_1^{tot}$ the part of the curve accessible is concave. In this condition, it is easy to see that the configuration of the cells that maximizes information transmission is configuration 2 ($S_1^A = S_1^B= S_1^{tot}/2$). 
At large values of $S_1^{tot}$, the part of the curve that is accessible is convex. In this condition, the configuration that maximizes information transmission is configuration 1 ($S_1^A \ne S_1^B$). 
}
\label{SupplFig_independent}
\end{figure}

\begin{figure}[ht]
    \centering
    \includegraphics[width=0.85\textwidth]{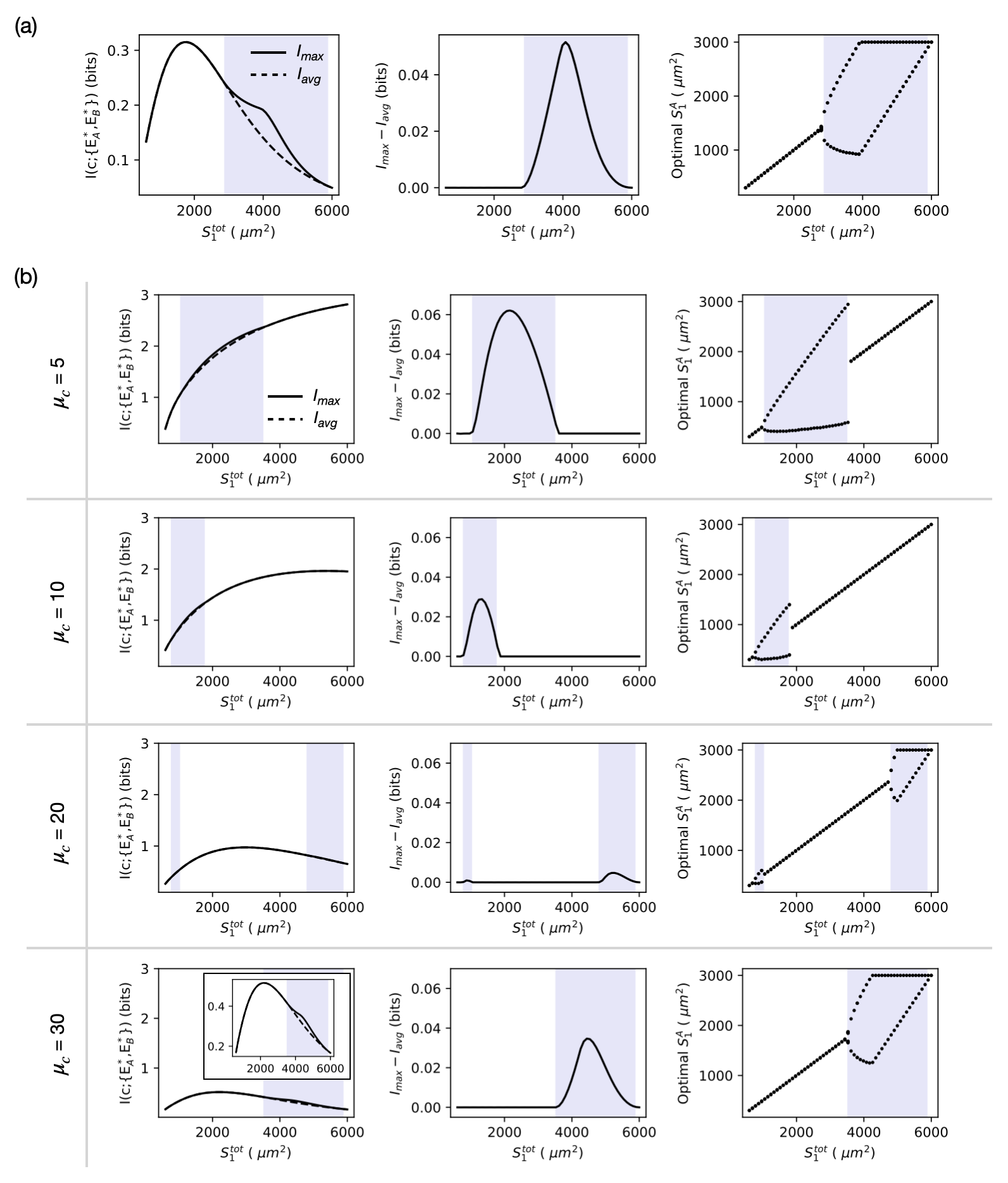}
    \caption{(a) Information transmission in the absence of ephrin. The plot on the left shows the maximal information ($I_{\rm max}$) transmitted between $c$ and $\{E^*_A, E^*_B\}$ (solid line, obtained when the cells expose the optimal surfaces to FGF) and the information transmitted when the two cells expose the same surface to FGF ($I_{\rm avg}$, dashed line) as a function of $S_1^{\rm tot}$. The plot in the middle shows the difference between $I_{\rm max}$ and $I_{\rm avg}$ as a function of $S_1^{\rm tot}$. The plot on the right shows the values of the optimal surface $S_1^{A*}$ as a function of  $S_1^{tot}$. The purple background highlights the regions of the plots where it is better, in order to maximize information transmission, for the two cells to break the symmetry and expose different surfaces to FGF. These plots are obtained assuming that the input distribution  $P(c)$ is centered around $\mu_c = 40$ and that the two cells have the same total surface $S_{\rm cell}^A = S_{\rm cell}^B = 6000 \ \mu m^2$. To model the absence of ephrin we used $e=10^{-5}$. 
(b) Influence of the position $\mu_c$ of the input distribution on information transmission. The plots in the first column show the maximal information ($I_{\rm max}$) transmitted between $c$ and $\{E^*_A, E^*_B\}$ (solid line, obtained when the cells expose the optimal surfaces to FGF) and the information transmitted when the two cells expose the same surface to FGF ($I_{\rm avg}$, dashed line) as a function of $S_1^{\rm tot}$. The plots in the second column show the difference between $I_{\rm max}$ and $I_{\rm avg}$ as a function of $S_1^{\rm tot}$. The plots in the third column show the values of the optimal surface $S_1^{A*}$ as a function of  $S_1^{\rm tot}$. The purple background highlights the regions of the plots where it is better, in order to maximize information transmission, for the two cells to break the symmetry and expose different surfaces to FGF. These plots are obtained assuming that the two cells have the same total surface $S_{\rm cell}^A = S_{\rm cell}^B = 6000 \ \mu m^2$.}
\label{SupplFig_two_cells}
\end{figure}

\begin{figure}
    \centering
    \includegraphics[width=1\textwidth]{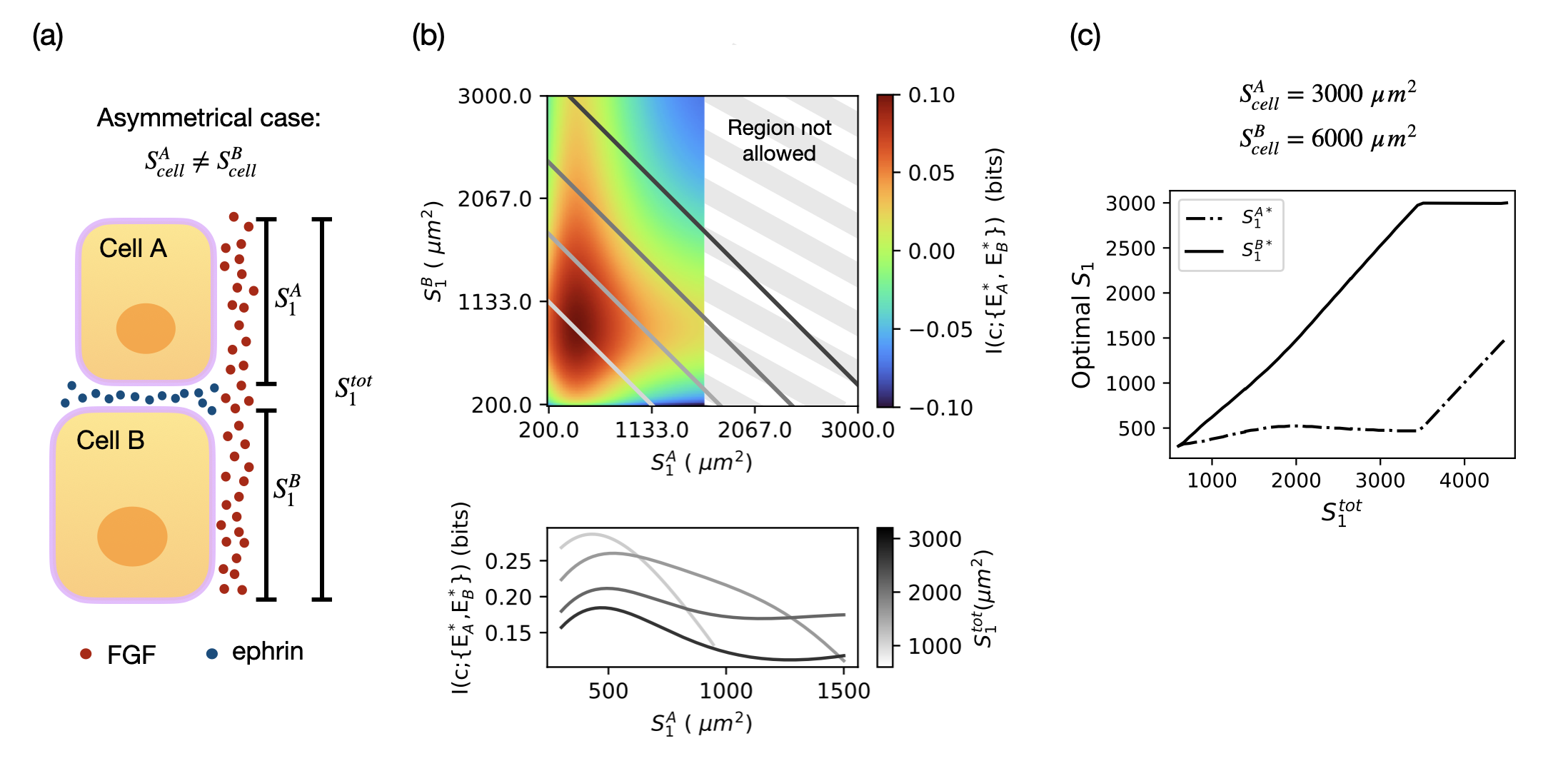}
    \caption{Two-cells: asymmetrical case.
    (a) In the asymmetrical case we consider that the two cells have different cell surface areas ($S_{cell}^A \ne S_{cell}^B$). $S_1^A$ and $S_1^B$ are the surface areas of the cells exposed to FGF.  $S_2^A$ and $S_2^B$ (not shown in the cartoon) are the surface areas of the cells exposed to ephrin. In each cell, $S_2^{A,B}$ is computed from the corresponding value of $S_1^{A,B}$ using equation Eq.\eqref{S1S2_relation}. The value of $S_1$ in each cell is constrained to be at most equal to $S_{\rm cell}/2$, with $S_{\rm cell}$ being the total surface area of the cell. $S_1^{\rm tot}$ represents the total surface area where FGF is present. The presence of the constraint implies that $S_1^A + S_1^B = S_1^{\rm tot}$. 
    (b) The heatmap represents the information transmitted between the FGF input $c$ and the ERK output $\{E^*_A, E^*_B\}$ as a function of the area of cell surfaces exposed to FGF ($S_1^A$ and $S_1^B$). The presence of the constraint restricts the possible values of $S_1^A$ and $S_1^B$ to straight lines defined by $S_1^B = S_1^{\rm tot} -S_1^A$. The straight lines colored with different shades of gray represent the accessible values of $S_1^A$ and $S_1^B$ for the values of $S_1^{tot}$ used in the plot below.
The plot below represents $I(c; \{ E_A^*, E_B^*\})$ as a function of $S_1^A$ for different values of the constraint  $S_1^{\rm tot}$ (see colorbar). 
(c) The plot shows the optimal values of the surfaces exposed to FGF ($S_1^{A*}$ with a solid line and $S_1^{B*}$ with a dashed line) as a function of the value of the constraint $S_1^{tot}$.  
The optimal values are obtained maximizing information transmission assuming $\mu_c = 40$, $S_{cell}^A = 3000 \ \mu m^2, S_{cell}^B =6000 \ \mu m^2$.}
    \label{fig:2cells_asymmetrical}
\end{figure}

\end{document}